\documentclass[fleqn,usenatbib]{mnras}

\usepackage{mathptmx}

\usepackage[T1]{fontenc}
\usepackage{ae,aecompl}

\usepackage{graphicx}	
\usepackage{amsmath}	
\usepackage{amssymb}	
\usepackage{natbib}
\usepackage{bm}			

\newcommand{\nab}{\mbox{\boldmath $\nabla$}}

\title[Coupling between fast tides and convection]{On a new formulation for energy transfer between convection and fast tides with application to giant planets and solar type stars}

\author[Caroline Terquem]{
Caroline Terquem\thanks{caroline.terquem@physics.ox.ac.uk}
\\
Department of Physics, Oxford University, Keble Road, Oxford OX1 3RH, UK\\
Institut d'Astrophysique de Paris, Sorbonne Universit\'e, CNRS,
  UMR 7095, 98 bis boulevard Arago, F-75014, Paris, France \\
}

\date{Accepted XXX. Received YYY; in original form ZZZ}

\pubyear{2020}

\begin{document}
\label{firstpage}
\pagerange{\pageref{firstpage}--\pageref{lastpage}}
\maketitle

\begin{abstract}
  All the studies of the interaction between tides and a
  convective flow assume that the large scale tides can be described
  as a mean shear flow which is damped by small scale fluctuating
  convective eddies.  
  The convective Reynolds stress is calculated using mixing length
  theory, accounting for a sharp suppression of dissipation when the
  turnover timescale  is larger than the tidal period.
  This yields   tidal dissipation rates several orders of magnitude too small to account for the circularization periods of late--type
  binaries or the tidal dissipation factor  of giant planets.  Here, we argue that the above description is
  inconsistent, because fluctuations and mean flow should be identified
  based on the timescale, not on the spatial scale, on which they
  vary.  Therefore, the standard picture should be reversed, with the fluctuations being the tidal oscillations and the mean shear flow provided by the largest convective eddies.    We assume that energy is locally transferred from the tides to the
convective flow.  Using this assumption,  we obtain values for the tidal $Q$ factor of Jupiter and  Saturn  and for the circularization periods of PMS binaries in good agreement with observations.  The timescales obtained with  the equilibrium tide approximation  are however still 40 times too large to account for the circularization periods of late--type binaries.   For these systems, shear in the tachocline or at the base of the convective zone  may be the main cause of tidal dissipation.  
\end{abstract}



\section{Introduction}

Tidal dissipation in stars and giant planets plays a very important role in shaping the orbits of binary systems.    
For early--type stars, which have a radiative envelope, tides are damped in the radiative surface layers.  The theory has been very successful at explaining the circularization periods of these stars \citep{Zahn1977}.
For late--type stars and giant planets, dissipation in the convective regions is expected to be  very important, although dissipation due to wave breaking in stably--stratified layers may also play a role \citep{Barker2010}.  In convective zones, the standard theory describes the tides as a mean flow which interacts with fluctuating convective eddies \citep{Zahn1966}.  The rate of energy transfer between the tides and the convective flow is given by the coupling between the Reynolds stress associated with the fluctuating velocities and the mean shear flow.  In this approach,  it is further argued that the fluctuations  vary on a  small enough spatial scale to justify the use of a diffusion approximation  to evaluate the Reynolds stress, leading to the introduction of a `turbulent viscosity' given by  mixing length theory.    In most cases of interest, the tidal periods are significantly smaller than the convective turnover timescale in at least part of the envelope.   In such a situation, convective eddies cannot transport and exchange momentum with their environment  during a tidal period, and dissipation is suppressed.  
  Rather than motivate a revision of the basic structure of the model,  this has been taken into account by incorporating a period--dependent term in the expression for the turbulent viscosity  (\citealt{Zahn1966}; \citealt{Goldreich1977}).   Tidal dissipation calculated this way is orders of magnitude too small to account for either the circularization period of late--type binaries, or the tidal dissipation factor of Jupiter and Saturn  inferred from the orbital motion of their  satellites.  This is still the  case even when the correction to the turbulent viscosity for large turnover timescales is formally ignored, or when resonances with dynamical  tides are included (\citealt{Goodman1997}; \citealt{Terquem1998}; \citealt{Ogilvie2014} and references therein).  
  
 Numerical simulations have attempted to measure the turbulent viscosity and its period dependence in local models \citep{Penev2009, Ogilvie2012, Duguid2020}, and the first global simulations have been published very recently \citep{Vidal2020a, Vidal2020b}.  Interestingly, the simulations (in the four more recent publications) show that the turbulent viscosity actually  becomes negative at large forcing frequencies.    This suggests that the standard picture of convective  turbulence  dissipating the tides is dubious when the period of the tides is smaller than the turnover timescale, even though negative viscosities are only obtained for unrealistically low tidal periods  \citep{Duguid2020, Vidal2020b}.  
 
 In this paper, we revisit the interaction between tides and convection in this regime.  In section~\ref{sec:conservation}, we show that, when the timescales can be well separated, traditional roles are reversed: the Reynolds decomposition yields energy equations in which the tides are the fluctuations,  whereas convection is the mean flow.  The spatial scales on which these flows vary is not relevant in identifying the fluctuations  and the mean flow.   In section~\ref{sec:transfer},  assuming  equilibrium tides, we give an expression for the rate $D_R$ at which the Reynolds stress exchanges energy  between the  tides and the convective flow.  Although  the sign of $D_R$ is not known, we make the strong assumption that energy is locally transferred from the tides to the convective flow ($D_R>0$), and investigate whether such a coupling yields an energy dissipation at the level needed to account for observations.  In section~\ref{sec:timescales}, we give expressions for the total dissipation rate corresponding to both circular and eccentric orbits, and for the orbital decay, spin up and circularization timescales.  We apply those results in section~\ref{sec:applications}.  We calculate the tidal dissipation $Q$ factor for Jupiter and Saturn, the circularization periods of pre--main sequence (PMS)  and late--type binaries and evolution timescales for hot Jupiters.  Apart from the notable exception of the circularization periods of late--type binaries,  all these results are in  good quantitative agreement with observations.  In section~\ref{sec:Discussion}, we discuss our results.  We also review  numerical simulations and observations of the Sun, which show that the interaction between convection and rotation leads to large scale flows and structures which are quite different from the traditional picture, and may produce the convective velocity gradients required to make $D_R>0$. 

\section{Conservation of energy in a convective flow subject to a fast varying tide} 
\label{sec:conservation}

We consider a binary system made of two late--type stars which orbit each other with a period $P_{\rm orb}$.  The period of the tidal oscillations excited in each of the stars by their companion, which  is $P=P_{\rm orb}/2$ for non--rotating stars,  is on the order of a few days for close binaries.
 We can estimate the convective turnover timescale $t_{\rm conv}$ in the convective envelope of the stars by assuming that all the energy is transported by convection.  The largest eddies cross the convective envelope on a time of order $t_{\rm conv}$, transporting the  kinetic energy of order $M_{\rm env} V^2$, where $V$ is the velocity of the eddies and $M_{\rm env}$ is the mass of the convective envelope.  The luminosity of the star is therefore $L \sim M_{\rm env} V^2 / t_{\rm conv}$.  To within a factor of order unity, $V \sim R/ t_{\rm conv}$, where $R$ is the radius of the star.  This yields $t_{\rm conv} \sim 40$~days for the Sun, which is significantly larger than $P$.  More precise solar models confirm that the  convective turnover  timescale is larger than a few days in a large part of the envelope. This  timescale can be interpreted as the lifetime of the convective eddies.  Therefore, the timescale $P$ on which the velocity of the fluid elements induced by tidal forcing varies  is much smaller than the  timescale $t_{\rm conv}$ on which the velocity of the largest convective eddies induced by buoyancy varies.   

\subsection{Reynolds decomposition and exchange of energy between the tides and convection}

We now consider a simplified model in which a flow is the superposition of two flows  which vary with very different timescales $\tau_1$ and $\tau_2 \gg \tau_1$, and outline for clarity the derivation of the standard equations which govern the evolution of the kinetic energy of the two flows, as this is at the heart of the argument we present in this paper (see, e.g.,  \citealt{Tennekes1972}  for details).     Compressibility is not important for the argument, so we assume that the flow is incompres\-sible (the analysis done in this section will be applied to equilibrium tides, which correspond to  incompressible fluid motions).   We use the Reynolds decomposition in which the total velocity $ {\bf u}$ is written as the sum of the velocity ${\bf V} $ of the slowly varying flow and that ${\bf u}'$ of the rapidly varying flow:
\begin{equation}
  {\bf u} = {\bf V} + {\bf u}',
\end{equation}
where ${\bf V} =\left< {\bf u} \right>$ and
$\left< {\bf u}' \right> = {\bf 0}$, with the brackets denoting an
average over a time $T$ such that  $\tau_	1 \ll T \ll \tau_2$.    A similar
decomposition can be made for the pressure $p$ and the viscous stress
tensor $\sigma_{ij}$: 
\begin{equation}
  p = P + p', \; \; \;  \sigma_{ij} = S_{ij}+ \sigma'_{ij},
\end{equation}
where:
\begin{equation}
\sigma_{ij} = \rho \nu \left( \frac{\partial u_i}{\partial x_j} +
  \frac{\partial u_j}{\partial x_i} \right),
\label{sigmaij2}
\end{equation}
with $\nu$ being the (molecular) kinematic viscosity, and
$P=\left< p \right>$, $S_{ij}=\left< \sigma_{ij} \right>$,
$\left< p' \right> = \left< \sigma'_{ij} \right> =0$.
The indices $i$ and $j$ refer to  Cartesian coordinates.  Molecular viscosity is not important for the dissipation of tides, but we keep this term as it helps to interpret the energy conservation equations.  
Incompres\-sibility implies:
\begin{equation}
  \nab \cdot \left( {\bf V} + {\bf u}'     \right) =0.
  \label{eq:incompturb}
\end{equation}
Taking a time--average of this equation yields:
\begin{equation}
  \nab \cdot {\bf V}    =0.
  \label{eq:incave}
\end{equation}
Subtracting
from equation~(\ref{eq:incompturb}) then gives:
\begin{equation}
  \nab \cdot {\bf u}'  =0,
  \label{eq:incfluc}
\end{equation}
which means that both the average flow and the  fluctuations are  incompressible.   We also assume that $\rho$ is constant with time and uniform.  Although this model is of course not a realistic description of the convective flow in a star, it contains the key ingredients for the argument which is presented here.  

The flow satisfies Navier--Stokes equation, which $i$--component  is:
\begin{equation}
  \frac{\partial u_i}{\partial
  t} +   \left( {\bf u} \cdot \nab \right) u_i = - \frac{1}{\rho} \frac{\partial
  p}{\partial x_i} +
\frac{1}{\rho}  \frac{\partial \sigma_{ij}}{\partial x_j} + \frac{1}{\rho} f_i,
\label{eq:NSturb}
\end{equation}
where ${\bf f}$ includes all the forces per unit volume which act on the
fluid, and we adopt the convention that repeated indices are summed over.  Substituting the Reynolds decomposition above
and  averaging the equation over the time  $T$ yields: 
\begin{equation}
  \frac{\partial V_i }{\partial
    t} +  \left( {\bf V} \cdot \nab \right) 
      V_i 
  +  \left< \left(   {\bf u}' \cdot \nab \right)  u'_i \right>=
   - \frac{1}{\rho} \frac{\partial  P}{\partial x_i} +
    \frac{1}{\rho} \frac{\partial S_{ij}}{\partial x_j}  +  \frac{1}{\rho} f_i ,
   \label{eq:NSturb1}
\end{equation}
where we have used the fact that the time and space
derivatives can be interchanged with the averages (for the time derivative, this is because $\tau_1 \ll T \ll \tau_2$).

The average kinetic energy per unit mass is:
$$ \left< \frac{1}{2} u_i u_i \right> =  \left< \frac{1}{2} \left( V_i + u'_i \right)\left( V_i + u'_i \right)  \right> =
 \frac{1}{2} \left( V_iV_i + \left<  u'_i u'_i \right>
\right),$$ which is the sum of the kinetic energy of the mean flow and
that of the fluctuations.   
\\

We obtain an energy conservation equation for the mean flow by multiplying
equation~(\ref{eq:NSturb1}) by $V_i$.  Using equations~(\ref{eq:incave}) and~(\ref{eq:incfluc}) then yields:
 \begin{multline}
  \frac{\partial  }{\partial
    t} \left( \frac{V_i V_i}{2} \right) +  V_j
  \frac{\partial  }{\partial x_j}  \left( \frac{V_i V_i}{2}  \right) = \\
    \frac{\partial}{\partial x_j}
  \left(
- \frac{V_j P}{\rho} + \nu V_i  \frac{\partial V_i 
  }{\partial x_j}  -V_i \left<  u'_i  u'_j 
   \right>
\right) +\frac{1}{\rho} V_i f_i  
-\nu  \frac{\partial V_i
  }{\partial x_j}  \frac{\partial V_i
  }{\partial x_j}   + D_R,
   \label{eq:NSturbE2}
   \end{multline}
where we have defined:
\begin{equation}
 D_R \equiv \left<  u'_i  u'_j 
\right> \frac{ \partial V_i }{ \partial x_j }.  
\label{eq:DR}
\end{equation}
This equation indicates  that the
Lagrangian derivative of the kinetic energy of the mean flow per unit mass (left hand--side) is equal to the
divergence of a
flux, which represents the work done by pressure forces, 
viscous and  Reynolds stresses on the mean flow,  plus the work done on the mean flow by the forces which act on the volume of the fluid, plus a term expressing dissipation of
energy in the mean flow due to 
viscosity, plus the term $D_R$, which represents the rate at which energy is  fed into or extracted from the mean flow by the Reynolds stress $R_{ij}=-\rho \left<  u'_i  u'_j 
\right>$.  

A similar conservation equation for the fluctuations can be obtained
by multiplying equation~(\ref{eq:NSturb}) by $u'_i$.  Substituting the Reynolds decomposition, averaging over time and using equations~(\ref{eq:incave}) and~(\ref{eq:incfluc}) then yields:
\begin{multline}
   \frac{\partial  }{\partial
    t} \left( \frac{\left< u'_i u'_i \right>}{2} \right) +  V_j
  \frac{\partial  }{\partial x_j}  \left( \frac{\left< u'_i u'_i
      \right>}{2}  \right) = \\
    \frac{\partial}{\partial x_j}
  \left(
- \frac{\left< u'_j p' \right>}{\rho} + \nu \left< u'_i
  \frac{\partial u'_i 
  }{\partial x_j} \right>  - \frac{\left<  u'_i  u'_i u'_j 
   \right>}{2}
\right) +\frac{1}{\rho} u'_i f_i  \\
-\nu  \left< \frac{\partial u'_i
    }{\partial x_j}   \frac{\partial u'_i
    }{\partial x_j}   \right> - D_R.
   \label{eq:NSturbE5}
   \end{multline}
   Here again, this equation indicates that the Lagrangian
   derivative of the kinetic energy of the fluctuations per unit mass  (left hand--side) is equal
   to the divergence of a flux, which represents the average of the
   work done by the fluctuating pressure forces, viscous and Reynolds
   stresses on the fluctuations, plus the work done on the fluctuations by the forces which act on the volume of the fluid,  plus a term expressing dissipation of energy in the
   fluctuations due to viscosity,  minus the same $D_R$ term as in equation~(\ref{eq:NSturbE2}).

As can be seen from equations~(\ref{eq:NSturbE2}) and~(\ref{eq:NSturbE5}), $D_R$ represents the rate of energy per unit mass which is exchanged between the mean flow and the fluctuations {\em via} the Reynolds stress: when $D_R<0$, energy is transferred from the mean flow to the fluctuations whereas, when $D_R>0$, energy is transferred from the fluctuations to the mean flow.    

\subsection{Comparison with previous work}
\label{sec:comparison}

All the studies that have been done to date on the interaction between tides and convective flows  have relied on a description where the fluctuations are identified with the convective flow, whereas the mean flow is  identified with the tidal oscillations.  It is then assumed that energy is transferred from the tides to the convective eddies, in much the same way that energy is transferred from the mean shear to the turbulent eddies in a standard turbulent shear flow.  This is described using a turbulent viscosity, which is assumed to be a valid concept because  the mean flow is perceived to vary on large scales, whereas the fluctuations are viewed as varying on small scales. 

In his pioneering study of tides in stars with convective envelopes, \citet{Zahn1966} assumed that convection could be described using a turbulent viscosity, which yields a viscous force acting on tidal oscillations.   He recognized that dissipation was reduced when the period $P$ of the oscillations was smaller than the convective turnover timescale $t_{\rm conv}$, and proposed a reduction by a factor $P/t_{\rm conv}$ in this context.   In \citet{Zahn1989}, he further commented that the concept of a turbulent viscosity relies on a diffusion approximation, only valid when the convective eddies vary on a spatial scale much smaller than that associated with the tides.    In a seminal paper, \citet{Goldreich1966} derived constraints on  tidal dissipation in planets in the solar system based on the evolution of their satellites.  They further estimated the amount of dissipation in Jupiter by assuming that damping of the tides occurred in a turbulent  boundary layer at the bottom of the atmosphere, where a solid core is present.   Later, \citet{Hubbard1974} investigated tidal dissipation in Jupiter assuming the existence of a viscosity in the interior of the planet.  He estimated its value using the constraints derived by \citet{Goldreich1966}, and concluded that the likely origin of this viscosity was turbulent convection.  His calculation did not take into account a reduction of dissipation for $P/t_{\rm conv}<1$.  \citet{Goldreich1977} subsequently pointed out that \citet{Hubbard1974} had overestimated tidal dissipation, and proposed a reduction of the turbulent viscosity by a factor $\left( P/t_{\rm conv} \right)^2$ in the regime $P/t_{\rm conv}<1$.  Neither  \citet{Hubbard1974} nor \citet{Goldreich1977}  referred to \citet{Zahn1966},  which indicates that they were not aware of his earlier work.  This may be because Zahn's 1966 papers were  written in French.  Following these earlier studies, there has been  much discussion about the factor by which  turbulent viscosity is reduced when $P/t_{\rm conv}<1$, but it has always been assumed that, in this regime,  convection could still be described as a turbulent viscosity damping the tides.  As already pointed out, this implicitly assumes that the spatial scales associated with convection are much smaller than that associated with the tides. 

As we will see below, the assumption that the tides vary on a scale larger than the largest  convective eddies is not always justified.  But, even more importantly,   equations~(\ref{eq:NSturbE2}) and~(\ref{eq:NSturbE5}) are obtained by identifying and separating  the mean flow and the fluctuations   based solely on the {\em timescales} on which they vary,  not on the spatial scales.    Therefore,  in the case of fast tides ($\tau_1=P$) interacting with slowly varying convection ($\tau_2=t_{\rm conv}$),  the fluctuations are  the tidal oscillations and the mean shear flow is provided by the largest convective eddies.  This implies that {\em the Reynolds stress $- \rho \left<  u'_i  u'_j 
\right>$ is given by the correlations between the components of the velocity of the tides, not that of the convective velocity}.  It is the coupling of this stress to the mean shear associated with the convective velocity which controls the exchange of energy between the tides and the convective flow.     

As far as we are aware, the term $D_R$ given by equation~(\ref{eq:DR}) has never been included in previous studies of tidal dissipation in convective bodies.  This term, however, is present in the energy conservation equation for the fluctuations even when a linear analysis of the tides is carried out, as it comes from the $u'_j \left( \partial V_i / \partial x_j \right)$ term in Navier--Stokes equation.  In \citet{Goodman1997}, it is eliminated on the assumption that it does not contribute to dissipation and, in \citet{Ogilvie2012} and \citet{Duguid2020}, it cancels out for the particular form of the flow chosen to model the tides. 

\section{Transfer of energy between the tides and the large convective eddies}
\label{sec:transfer}

In the case of a standard turbulent shear flow, the Reynolds stress is given by the correlations between the components of the turbulent velocity, and the coupling to the background mean shear determines how energy is exchanged.  Because the length--scale of the turbulent eddies is  small compared to the scale of the shear flow,  eddies are stretched by the shear flow, and conservation of angular momentum then produces a correlation of the components of the turbulent velocity yielding $D_R<0$ (see, e.g., \citealt{Tennekes1972}).  This corresponds to a transfer of energy from the mean flow to the largest turbulent eddies and the subsequent cascade results in a small scale viscous dissipation of the free energy  present in the shear flow.  

In the case of fast tides interacting with slowly varying convection, fluid elements oscillating because of the tidal forcing cannot be stretched by the mean flow associated with convection in the same way as described above, because the length--scale of the tides may be larger, sometimes even much larger,  than that of the eddies and also because the tides are imposed by an external forcing.  Therefore, in this context, there is no reason why  energy would be transferred from the mean convective flow to the tides, which would correspond to $D_R<0$.  In addition, 
 if $D_R$ were negative,  the  amplitude of the tides would be increased by the interaction with convection, which in turn would increase the orbital eccentricity   of the binary (see  \citealt{Goldreich1966} for a physical explanation of how tidal interaction modifies the eccentricity of the orbit).   Also, this would lead to a decrease of the orbital period when the rotational velocity of the body in which the tides are raised is larger than the orbital velocity of the companion.
This would not be in agreement with observations, which indicate that tides are dissipated when interacting with a  convective flow: this is evidenced by the circularization of late--type binaries and the orbital evolution of the satellites of Jupiter and Saturn.  
This implies that  there is a net transfer of kinetic energy from the tides to the convective eddies, that is to say 
the integral of $\rho D_R$, where $\rho$ is the mass density, over the volume of the convective zone is positive.  Equations~(\ref{eq:NSturbE2}) and~(\ref{eq:NSturbE5}) have been obtained by averaging the motion over a time $T$ which is small compared to the timescale $\tau_2=t_{\rm conv}$ over which the convective eddies vary, which amounts to considering they are `frozen'.  Therefore, these equations cannot be used to understand how energy is transferred from the tides to the convective eddies.  If fast tides always transfer kinetic energy to the largest convective eddies, there has to be some universal mechanism by which the flow re--arranges itself to make the integral of $\rho D_R$ positive.   In the envelope of the Sun, convection interacting with rotation
 does not look like the standard picture of blobs going up and down.  In particular, the Coriolis force inhibits radial downdrafts near the equator, and rotation produces prominent columnar structures, as expected from the Taylor--Proudman theorem \citep{Featherstone2015}.   
 This will be discussed further in section~\ref{sec:Discussion}.
 Calculating $D_R$ requires knowing the gradient of the convective velocity which, as of today, cannot be obtained  even  from state--of--the--art numerical simulations.   Therefore,  in order to progress, we have to  make very crude assumptions and approximations.  
   Thereafter,  we will then assume that the gradient of the convective velocity is such that $D_R$ is everywhere positive in convective regions.  
 The idea is to investigate whether the maximum energy dissipation obtained in that ideal case  would be at the level needed to explain the circularization period of late--type binaries and the tidal dissipation factor of Jupiter and Saturn.  Note that, although this is a very strong assumption,  it is similar to the assumption made in all previous studies that the turbulent Reynolds stress associated with convection couples positively to the gradient of the tidal velocities to extract energy from the tides.  

 We now evaluate the correlation of the components of the tidal velocity, $\left< u'_i u'_j \right>$, assuming   equilibrium tides (which satisfy the assumption of incompressible
  fluid motions  made in the analysis of section~\ref{sec:conservation}).   
 The equilibrium tide approximation is  actually rather poor  in convective regions where the  Brunt--V\"ais\"al\"a frequency   is not very large compared to the tidal frequency,  and this yields to an over--estimate of  tidal dissipation by a factor of a few for close binaries \citep{Terquem1998,  Barker2020}. 
 It also does not apply 
  in a thin region near the surface of the convective envelope \citep{Bunting2019}.  However, given all the uncertainties in estimating tidal dissipation here,  the equilibrium tide approximation is sufficient.   To zeroth order in eccentricity and for a non--rotating body, this gives ${\bf u}' = \partial  {\bm \xi} / \partial t $ with (e.g., \citealt{Terquem1998}):
\begin{align}
\xi_r \left( r, \theta, \varphi , t \right) & =  f \xi_r (r) \times 3 \sin^2 \theta \cos  \left( m \varphi -  n \omega_{\rm orb} t \right)  , 
\label{eq:xir} \\
\xi_{\theta}  \left( r, \theta, \varphi , t \right) & =  f \xi_h(r)   \times 6 \sin \theta \cos \theta \cos  \left( m \varphi - n \omega_{\rm orb} t \right) , 
\label{eq:xitheta} \\
\xi_{\varphi}   \left( r, \theta, \varphi , t \right) & =  - f \xi_h(r)  \times 3m \sin \theta \sin   \left( m \varphi - n \omega_{\rm orb} t \right)  , 
\label{eq:xiphi}
\end{align} 
where:
\begin{align}
\xi_r (r) & = r^2 \rho \left( \frac{ {\rm d} P}{{\rm d} r} \right)^{-1}, 
\label{eq:xirr} \\
\xi_h (r) & = \frac{1}{6r} \frac{ {\rm d}}{{\rm d} r} \left( r^2 \xi_r (r) \right) , 
\label{eq:xihr} 
\end{align}
and $n=m=2$.  Here, $\omega_{\rm orb} $ is the orbital frequency and $f=- GM_p / 4 a^3$, with $M_p$ being the mass of the companion  which excites the tides, $a$ being the binary separation and $G$ being the gravitational constant.     The frequency of the tidal oscillation is $\omega=n \omega_{\rm orb}$, while the period is $P=P_{\rm orb}/n$, with $P_{\rm orb}=2 \pi / \omega_{\rm orb}$ being the orbital period.   
Using the equation of hydrostatic equilibrium, equation~(\ref{eq:xirr}) yields $\xi_r(r)=-r^4/ \left[ GM(r) \right]$, where $M(r)$ is the mass contained within the sphere of radius $r$.  Therefore, if $M(r)$ varies slowly with radius, as in the convective envelope of the Sun for example, $\xi_h(r) \simeq \xi_r(r)$. \\
This equilibrium tide  is the response of the star obtained ignoring convection and any other form of dissipation.  To calculate tidal dissipation in a self--consistent way, we  should in principle solve the full equations including  convection, and this would in particular introduce a phase shift between the radial and horizontal parts of the tidal displacement.  However, as dissipation is expected to be small (i.e., the energy dissipated during a tidal cycle is small compared to the energy contained in the tides), first--order perturbation theory can be used.  This means that the tidal velocities can be calculated ignoring dissipation, which can then be estimated from these velocities. This is the approach used in \citet{Terquem1998}. \\
The expressions above imply that 
  $\left< u'_r u'_{\varphi} \right> = \left< u'_{\theta} u'_{\varphi} \right> = 0$ and, 
  since $\xi_r (r)$ varies on a scale comparable to $r$,  $\left| \left< u'^2_{\theta} \right> \right| \sim \left| \left< u'^2_{\varphi} \right> \right|
\sim \left| \left< u'^2_r \right> \right| \sim \left| \left< u'_r u'_{\theta} \right> \right| \sim u'^2$, where  $u'$  is the characteristic value of the tidal velocity. 
  Therefore, from 
  equation~(\ref{eq:DRspher}), which gives $D_R$ in spherical coordinates, we obtain:
  \begin{equation}
  D_R \sim u'^2 \frac{V}{H_c},
  \end{equation}
where  $V$ is the characteristic  value of the convective velocity  and $H_c$ is the scale over which it varies.   In standard studies of tides interacting with convection, it is assumed that the fluctuations are associated with the convective flow whereas the mean flow is the tidal oscillation.    In this picture,  dissipation by large eddies, with a long turnover timescale, is suppressed, which is accounted for by adding a period--dependent term to the dissipation rate per unit mass, which is then given by:
\begin{equation}
D^{\rm st}_R = \frac{ \left< V_i V_j \right> }{ 1+ \left(  t_{\rm conv} / P \right)^s } \frac{\partial u'_i }{\partial x_j} ,
\label{eq:DRst}
\end{equation}
where the superscript `st' indicates that this dissipation rate corresponds to the standard approach. The value of $s=1$ was originally proposed by \citet{Zahn1966}, but it was later argued by \citet{Goldreich1977} that $s=2$ should be used instead (see \citealt{Goodman1997} for a clear presentation of the arguments). 
Mixing length theory is then used to calculate the Reynolds stress, which gives:
\begin{equation}
 \left| \left< V_i V_j \right> \right| \sim \frac{ \nu_t u'}{r},
 \label{eq:convstress}
 \end{equation}
 where 
$\nu_t \sim H_c V$ is the turbulent viscosity.    The new dissipation rate we propose can be compared to the standard value:
\begin{equation}
\frac{D_R}{D^{\rm st}_R} \sim \left( \frac{r}{H_c} \right)^2 \left[ 1+ \left( \frac{ t_{\rm conv} }{ P} \right)^s \right].
\label{eq:DRtide}
\end{equation}
If $r/H_c \gg 1$ and/or $t_{\rm conv} \gg P$, then  $D_R \gg D^{\rm st}_R$.

\section{Total dissipation rate in stars and giant planets and evolution timescales}
\label{sec:timescales}

The dissipation rate per unit mass in spherical coordinates is given by equation~(\ref{eq:DRspher}).  This equation shows that, in addition to $\left< u'_r u'_{\theta} \right>$, the quantities $\left< u'^2_r \right>$, $\left< u'^2_{\theta} \right>$ and $\left< u'^2_{\varphi} \right>$  may contribute to $D_R$.  The corresponding  terms in $D_R$ would add up to zero if the tides were completely isotropic and convection incompressible.  Although we have assumed in the analysis above that  convection was incompressible, it is not the case in reality, and as all  these terms may contribute we will retain them.   Of course the analysis is not consistent, since extra terms would have to be included in the energy conservation equation for compressible convection.  However, our conclusions do not depend on whether we include $\left< u'^2_r \right>$, $\left< u'^2_{\theta} \right>$ and $\left< u'^2_{\varphi} \right>$  or not, as we will justify below.  Equation~(\ref{eq:DRspher}) shows that $\left< u'^2_r  \right>$ couples to $V/ H_c$,   whereas $\left< u'^2_{\theta} \right>$ and $\left< u'^2_{\varphi} \right>$ couple to $V/r$.   For $\left< u'_r u'_{\theta} \right>$, the coupling is to both $V/H_c$ and $V/r$, with the dominant term being that associated with $V/H_c$ (as will be seen below, in the parts of the envelopes that contribute most to dissipation, $H_c < r$).  The dominant component of the convective velocity is usually taken to be in the $r$--direction, but as here we investigate the maximum dissipation rate that could be obtained, we allow for the possibility that horizontal components may play a role as well.  

Therefore, 
we approximate $D_R$ as:
\begin{equation}
D_R= \left( \left| \left< u'_r u'_{\theta} \right> \right| + \left|  \left< u'^2_r  \right> \right|  \right) \frac{V}{H_c} + \left( \left| \left< u'^2_{\theta} \right> \right| +\left| \left< u'^2_{\varphi} \right> \right| \right) \frac{V}{r} ,
\label{eq:DRtide1}
\end{equation}
where we have assumed that $D_R$ is positive, as discussed above. 

If the body in which the tides are raised rotates synchronously with the orbit,  the companion does not exert a torque on the tides.  In that case,   if the orbit is circular,  the semi--major axis stays fixed.  However, if the orbit is eccentric, although there is no net torque associated with the tides, there is still  dissipation of energy.  This leads to a change of semi--major axis, which has to be accompanied by a change of eccentricity  $e$ to keep the orbital angular momentum constant.    For the parameters of interest here, $e$ always decreases \citep{Goldreich1966}.   

Therefore, energy dissipation in a synchronously rotating body requires the perturbing potential to be expanded to non--zero orders in  $e$.   Such an expansion is also needed to calculate the circularization timescale, whether the body is synchronous or not, 
as both zeroth and first order  terms in $e$ in the expansion of the potential contribute to this timescale at the same order (e.g., \citealt{Ogilvie2014}).  An expansion to first order in $e$ is sufficient, as higher order terms lead to short timescales and therefore a rapid decrease of $e$.  Most of the circularization process is therefore dominated by the stages where $e$ is small \citep{Hut1981, Leconte2010}.    We now calculate the total dissipation rate for both  circular and eccentric orbits, in the limit of small $e$.

\subsection{Dissipation rate for a circular orbit}

\label{sec:dissipcircular}

 The total rate of  energy dissipation in the convective envelope is:
\begin{equation}
\left( \frac{{\rm d}E}{{\rm d}t} \right)_{\rm c} =  2 \int_{\left( t_{\rm conv} > \frac{P_{\rm orb}}{n} \right) }  \int_0^{\pi/2} \rho D_R \times 2 \pi r^2 \sin \theta {\rm d} \theta {\rm d} r  ,
\label{eq:dEdttheta}
\end{equation}
where the subscript `c' indicates that the calculation applies to a circular orbit.  Using equations~(\ref{eq:DRtide1}), this yields:
\begin{equation}
\left( \frac{{\rm d}E}{{\rm d}t} \right)_{\rm c} =  \frac{6}{5} \pi   n^2 \omega^2_{\rm orb}  f^2 I_1 \left( \omega_{\rm orb}, m, n \right),
\label{eq:dEdt}
\end{equation}
with:
\begin{multline}
I_1 \left( \omega_{\rm orb}, m, n \right)= \int_{\left( t_{\rm conv} > P_{\rm orb}/{n} \right) } {\rm d} r \; \rho(r)  \times \\ \left\{ 
\left[ r \xi_r (r)  \frac{{\rm d}}{{\rm d}r} \left( r^2 \xi_r (r) \right) + 8 r^2 \xi^2_r(r) \right] \frac{V(r)}{H_c(r)} 
\right.
\\
\left. 
+ \frac{4 + 5m^2}{18}    \left[ \frac{{\rm d}}{{\rm d}r} \left( r^2 \xi_r (r) \right) \right]^2 \frac{V(r)}{r}
\right\}.
\label{eq:I1}
\end{multline}
Note that $I_1$ may depend on $\omega_{\rm orb}$ and $n$, as the domain of integration covers the region where $t_{\rm conv}>P=P_{\rm orb}/n$.    In principle, we should add the contribution arising from $D^{\rm st}_R$ over the domain where $t_{\rm conv} < P$.  However, this is very small compared to the integral above, as will be justified later, so it can be neglected.    

For a binary system where a body of mass $M_p$ raises tides on a  body of mass $M_c$, we can make the scaling of
$ {\rm d} E / {\rm d} t $ with $\omega_{\rm orb}$ and $M_p$ clear by using $f=-M_p \omega^2_{\rm orb}/ \left[ 4 \left( M_c+M_p \right) \right]$.  This yields:
\begin{equation}
\left( \frac{{\rm d}E}{{\rm d}t}  \right)_{\rm c} = \frac{3 n^2}{40} \pi \left(  \frac{M_p}{M_c+M_p} \right)^2 \omega^6_{\rm orb} I_1 \left( \omega_{\rm orb}, m, n \right) .
\label{eq:dEdtc}
\end{equation}
 For a fixed $n$, as $\omega_{\rm orb}$ increases, $P$ decreases and therefore $I_1$ may increase (it  happens if $t_{\rm conv} >P$  only in part of the envelope).  This implies that $\left( {\rm d}E / {\rm d}t \right)_{\rm c} \propto \omega^q_{\rm orb}$ with $q \ge 6$.  For comparison, 
\citet{Terquem1998} obtained $\left( {\rm d}E / {\rm d}t \right)_{\rm c}  \propto \omega^5_{\rm orb}$ using the standard model with turbulent viscosity.  

So far, we have considered a non--rotating body.  Calculating the response of a rotating body to a tidal perturbing potential is very complicated and beyond the scope of this paper.  
We can however make an argument to estimate how the rate of energy dissipation calculated above would be modified if  the body rotated.  In the simplest approximation where the body rotates rigidly with uniform angular  velocity $\Omega$,  the tides retain the same radial structure but  each component  rotates at a velocity  $n \omega_{\rm orb} - m \Omega$ in the frame of  the fluid (where, for a circular orbit, $n=m=2$).   A standard approach would be to use the above derivation of ${\rm d} E / {\rm d}t$ and shift the velocity of the tide accordingly (as done in \citealt{Savonije1984}).  However, what matters in calculating the dissipation rate $D_R$ in equation~(\ref{eq:DRtide1}) is not the velocity of the tide {\em relative} to the equilibrium fluid in the body, but the velocity of the tide and that of the convective flow in an inertial frame.  This suggests that the calculation of $D_R$ is roughly the same whether the body rotates or not.  However, the integral $I_1$ is calculated over the domain where $t_{\rm conv}$ is larger than the period of the tide, and this does involve the frequency of the tide {\em relative} to the fluid in the body.  This suggests that   the  energy dissipation rate when the body rotates  is still given by equation~(\ref{eq:dEdt}),  but with the appropriate modification for the domain of integration of $I_1$.  

\subsection{Dissipation rate for an eccentric orbit}

\label{sec:dissipeccentricity}

We now  calculate the energy dissipation rate  in the limit of small eccentricity following the method presented in \citet{Savonije1983}.    To first order in $e$,   and assuming a non--rotating body,  the perturbing potential can be written as:
\begin{equation}
\Phi_p =  f r^2 \left[  \Phi_{2,2} + e \left( \Phi_{0,1} +  \Phi_{2,1} +  \Phi_{2,3} \right) \right] ,
\end{equation} 
where the subscripts indicate the values of $m, n$.  We have \citep{Savonije1983, Ogilvie2014}:
\begin{align}
\Phi_{2,2} & = 3 \sin^2 \theta \cos  \left( 2 \varphi -  2 \omega_{\rm orb} t \right)  ,  \label{eq:phi22} \\
\Phi_{0,1} & = 3  \left( 3 \cos^2 \theta -1 \right) \cos  \left(  \omega_{\rm orb} t \right)  ,  \\
\Phi_{2,1} & = \frac{3  }{2} \sin^2 \theta \cos  \left( 2 \varphi -   \omega_{\rm orb} t \right)  ,  \\
\Phi_{2,3} & =-\frac{21 }{2} \sin^2 \theta \cos  \left( 2 \varphi -  3 \omega_{\rm orb} t \right)  . \label{eq:phi23}
\end{align} 
The tidal displacement corresponding to each component can be written as in equations~(\ref{eq:xir})--(\ref{eq:xiphi}) but with the appropriate angular and time dependence (which, for $\xi_{\theta}$ and $\xi_{\varphi}$, are obtained by applying $\partial / \partial \theta$ and $\partial / \left( \sin \theta \partial \varphi \right)$, respectively,  to the  angular and time dependence of $\xi_r$).  
It is straightforward to show that the terms $\Phi_{2,1} $ and $\Phi_{2,3} $ contribute an energy dissipation rate given by equation~(\ref{eq:dEdt}) with the appropriate value of $n$, but multiplied by $e^2/4$ and $49e^2/4$, respectively.  For  $\Phi_{0,1}$, the integral over $\theta$ in equation~(\ref{eq:dEdttheta}) has to be re--calculated, and this yields the same energy dissipation as given by equation~(\ref{eq:dEdt}) with $n=1$, but multiplied by $e^2$.  
Therefore, the total energy dissipation rate is:
\begin{multline}
\left( \frac{{\rm d}E}{{\rm d}t} \right)_{\rm e} =  \frac{6}{5} \pi    \omega^2_{\rm orb}  f^2 
 \bigg\{
 4 I_1 \left( \omega_{\rm orb},2, 2 \right) 
 \\ 
 + e^2 \left[  \frac{1}{4}  I_1 \left( \omega_{\rm orb}, 2, 1 \right) 
 + \frac{441}{4}  I_1 \left( \omega_{\rm orb}, 2, 3 \right) 
 +  I_1 \left( \omega_{\rm orb}, 0, 1 \right) 
 \right]
  \bigg\},
\label{eq:dEdtenr}
\end{multline}
where the terms in  braces correspond, in the order in which they appear, to the contributions from $\Phi_{2,2} $, $\Phi_{2,1} $, $\Phi_{2,3} $ and $\Phi_{0,1}$, respectively.  The subscript `e' indicates that the calculation applies to an eccentric orbit.  

 In some cases, the body is spun up and becomes synchronous before circularization is achieved.   As mentioned above,  when the body rotates, we expect the energy dissipation rate to be given by the same expression as for a non--rotating body, but with the domain of integration of $I_1$ to include the region where $t_{\rm conv}$ is larger than the period of the tide relative to that of the fluid.  
When the body is synchronized,  this amounts to replacing $I_1 \left( \omega_{\rm orb}, m, n \right)$ in equation~(\ref{eq:dEdtenr}) by $I_1 \left( \omega_{\rm orb}, m, \left| n-m \right| \right)$ for  the term contributed by  $\Phi_{m,n}$.  In addition, 
   the term due to $\Phi_{2,2}$ has to be removed as a circular orbit does not contribute to energy dissipation in that case.  We then obtain the following estimate for the rate of energy dissipation in a synchronized body:
\begin{multline}
\left( \frac{{\rm d}E}{{\rm d}t} \right)_{\rm e, sync} =  \frac{6}{5} \pi    \omega^2_{\rm orb}  f^2 e^2  \times \\
\left[  \frac{1}{4}  I_1 \left( \omega_{\rm orb}, 2, 1 \right) 
 + \frac{441}{4}  I_1 \left( \omega_{\rm orb},  2, 1 \right) 
 +  I_1 \left( \omega_{\rm orb}, 0, 1 \right) 
 \right],
 \label{eq:dEdtesync0}
\end{multline}
where the terms in braces correspond, in the order in which they appear, to the contributions from  $\Phi_{2,1} $, $\Phi_{2,3} $ and $\Phi_{0,1}$, respectively.  The subscript `e, sync' indicates that the calculation applies to an eccentric orbit and a synchronous body.   This can be written more simply as:
\begin{multline}
\left( \frac{{\rm d}E}{{\rm d}t} \right)_{\rm e, sync} =  \frac{3 \pi}{40}   \left(  \frac{M_p}{M_c+M_p} \right)^2  \omega^6_{\rm orb}   e^2  \times  \\
\left[  \frac{442}{4} I_1 \left( \omega_{\rm orb}, 2, 1 \right) + I_1 \left( \omega_{\rm orb}, 0, 1 \right) \right].
 \label{eq:dEdtesync}
\end{multline}

\subsection{Evolution timescales}

\subsubsection{Orbital decay}

The energy which is dissipated leads to a decrease of the orbital energy $E_{\rm orb}=-GM_c M_p / \left( 2 a \right)$,  such that ${\rm d}E_{\rm orb} / {\rm d}t = -{\rm d}E / {\rm d}t$, and therefore to a decrease of the binary separation.  The characteristic  orbital decay  timescale  is given by:
\begin{equation}
t_{\rm orb} \equiv - a \left( \frac{{\rm d} a }{{\rm d} t} \right) ^{-1} = 
 \frac{M_c M_p}{M_c + M_p} \frac{\omega^2_{\rm orb} a^2}{  2 \left( {\rm d}E / {\rm d}t \right)}.
 \label{eq:torb}
\end{equation}
If the body is synchronized, $ {\rm d}E / {\rm d}t $ is given by equation~(\ref{eq:dEdtesync}).  As $\left( d {\rm E} / {\rm d} t  \right)_{\rm e, sync} \propto e^2$, the timescale is very long for small eccentricities.  If the body is not synchronized,  the dominant contribution to the rate of energy dissipation comes from $\Phi_{2,2}$ for small eccentricities, and therefore $ {\rm d}E / {\rm d} t = \left( d {\rm E} / {\rm d} t  \right)_{\rm c} $ is given by equation~(\ref{eq:dEdt}).

\subsubsection{Spin up}

When the body of mass $M_c$ is non--rotating (or rotating with a period longer than  the orbital period), the companion exerts a positive torque $\Gamma $ on the tides which corresponds to a decrease of the orbital angular momentum.
An equal and opposite torque is exerted on the body of mass $M_c$, which angular velocity $\Omega$  therefore increases as $I \left( {\rm d} \Omega / {\rm d} t \right) = \Gamma$, where $I$ is the moment of inertia of the body.  
Assuming a circular orbit, we have  $\Gamma =  \left( {\rm d} E / {\rm d} t \right)_{\rm c} / \omega_{\rm orb} $,  which yields  the spin~up (or
 synchronization) timescale:
\begin{equation}
t_{\rm sp} \equiv  - \left(  \Omega - \omega_{\rm orb}   \right) \left( \frac{{\rm d} \Omega }{{\rm d} t} \right) ^{-1} \simeq 
\frac{I \omega^2_{\rm orb} }{    \left( {\rm d}E / {\rm d}t \right)_{\rm c} },
\label{eq:tsp}
\end{equation}
where we have used $\Omega \ll \omega_{\rm orb}$, as these are the values of $\Omega$ which contribute most  to $t_{\rm sync}$.

 \subsubsection{Circularization}

The rate of change of  eccentricity is obtained by writing the rate of change of orbital angular momentum $L_{\rm orb} $, where: 
\begin{equation}
L_{\rm orb} = \frac{M_c M_p}{M_c+M_p} \left[ G \left( M_c + M_p \right) a \left( 1-e^2 \right) \right]^{1/2}.
\label{eq:Lorb}
\end{equation}

A mentioned above, to calculate the circularization timescale, we need to expand the perturbing potential to first order in $e$.
For each of the components $\Phi_{m,n}$ of the potential given by equations~(\ref{eq:phi22})--(\ref{eq:phi23}), we calculate ${\rm d }L _{\rm orb} /{\rm d} t $ and  express ${\rm d}a/{\rm d}t$ as a function of ${\rm d} E/{\rm d}t$.  We then use the following relation (e.g., \citealt{Witte1999}):
\begin{equation}
n \omega_{\rm orb} \frac{{\rm d}L_{\rm orb}}{{\rm d}t}=m \frac{{\rm d}E_{\rm orb}}{{\rm d}t} = -m \frac{{\rm d}E}{{\rm d}t} ,
\label{eq:EdotHdot}
\end{equation}
to obtain:
\begin{equation}
\frac{M_p M_c}{M_p + M_c} \frac{\omega^2_{\rm orb} a^2 e^2}{1-e^2}  t^{-1}_{\rm circ} = 
\left( 1 - \frac{1}{\sqrt{1-e^2}} \frac{m}{n} \right) \frac{{\rm d}E}{{\rm d}t},
\end{equation}
where the circularization timescale is defined as:
\begin{equation}
t_{\rm circ}  = - e \left( \frac{{\rm d} e}{{\rm d}t} \right)^{-1}.
\end{equation}
Using the values of $ {\rm d} E / {\rm d} t$ contributed by each component of the potential, as written in equation~(\ref{eq:dEdtenr}),
we calculate $t^{-1}_{\rm circ}$  to zeroth order in $e$ for each of these components, and add all the contributions to obtain  $t^{-1}_{\rm circ}$ produced by the full potential.  This yields:
\begin{multline}
 \left( t ^{\rm nr}_{\rm circ} \right)^{-1}=  \frac{3 \pi}{10} \frac{M_p}{M_c +M_p}  \frac{\omega^4_{\rm orb}}{M_c a^2}   
\left[ -\frac{1}{2} I_1 \left( \omega_{\rm orb}, 2, 2 \right) \right.
\\ \left.
- \frac{1}{16}  I_1 \left( \omega_{\rm orb}, 2, 1 \right) + \frac{147}{16}  I_1 \left( \omega_{\rm orb},2, 3 \right) + \frac{1}{4}  I_1 \left( \omega_{\rm orb}, 0, 1 \right)  \right].
\label{eq:tcircnr}
\end{multline}
The terms in brackets correspond, in the order in which they appear, to the contributions from $\Phi_{2,2}$,  $\Phi_{2,1}$, $\Phi_{2,3}$ and  $\Phi_{0,1}$.  The superscript `nr' indicates that the calculation applies to a non--rotating body. 

If the body of mass $M_c$ rotates synchronously,   the  argument developed above suggests that the circularization timescale can be written in  the same way as  for a non--rotating star, but with $I_1 \left( \omega_{\rm orb}, m, n \right)$ in equation~(\ref{eq:tcircnr}) being replaced by $I_1 \left( \omega_{\rm orb}, m, \left| n-m \right| \right)$ for the term contributed by  $\Phi_{m,n}$.  Also, 
   the term contributed by  $\Phi_{2,2}$ should be removed for synchronous rotation.  We then obtain the following estimate for the circularization timescale: 
   \begin{multline}
 \left( t ^{\rm sync}_{\rm circ} \right)^{-1}=  \frac{3 \pi}{10} \frac{M_p}{M_c +M_p}  \frac{\omega^4_{\rm orb}}{M_c a^2}   \times \\
\left[   \frac{73}{8}  I_1 \left( \omega_{\rm orb},2, 1 \right) + \frac{1}{4}  I_1 \left( \omega_{\rm orb}, 0, 1 \right)  \right]
  .
\label{eq:tcircsync}
\end{multline}
The superscript `sync' indicates that the calculation applies to a synchronous body.

\section{Applications} 
\label{sec:applications}

We now apply these  results to Jupiter, Saturn, PMS and  late--type binaries and  systems with a star and a hot Jupiter.

\subsection{Jupiter's tidal dissipation factor}
\label{sec:Jupiter}

In this section, we evaluate the rate at which the energy of the tides raised by Io in Jupiter dissipates.  This corresponds to $P_{\rm orb}=42.5$~hours and $M_p=8.93 \times 10^{22}$~kg (Io's mass).    Jupiter's rotational period is 9.9~hours,  which is short compared to the orbital period, so that in principle the tides should be calculated taking into account rotation.   However, it has been found that Jupiter rotates as a rigid body, with differential rotation being limited to the upper 3,000~km, that is to say about 4\% of its atmosphere \citep{Guillot2018}, and the tidal response  taking into account solid body rotation is very well approximated by the equilibrium tide \citep{Ioannou1993}.  Interestingly, it has been found that Io is moving towards Jupiter \citep{Lainey2009}.   Tidal dissipation in Jupiter increases Io's   angular momentum and hence its orbital energy, since Jupiter's rotational velocity is larger than Io's orbital velocity.
However, the resonant interaction with the other Galilean satellites induces an orbital eccentricity which leads to tidal dissipation in Io itself (there would be no dissipation if the orbit were circular, as Io rotates synchronously with the orbital motion), decreasing its orbital energy.  The resonant interaction also directly decreases the orbital energy, and these losses  are larger than the gain from the exchange with Jupiter's rotation. 

\begin{figure*}
    \centering
   \includegraphics[width=\columnwidth,angle=270]{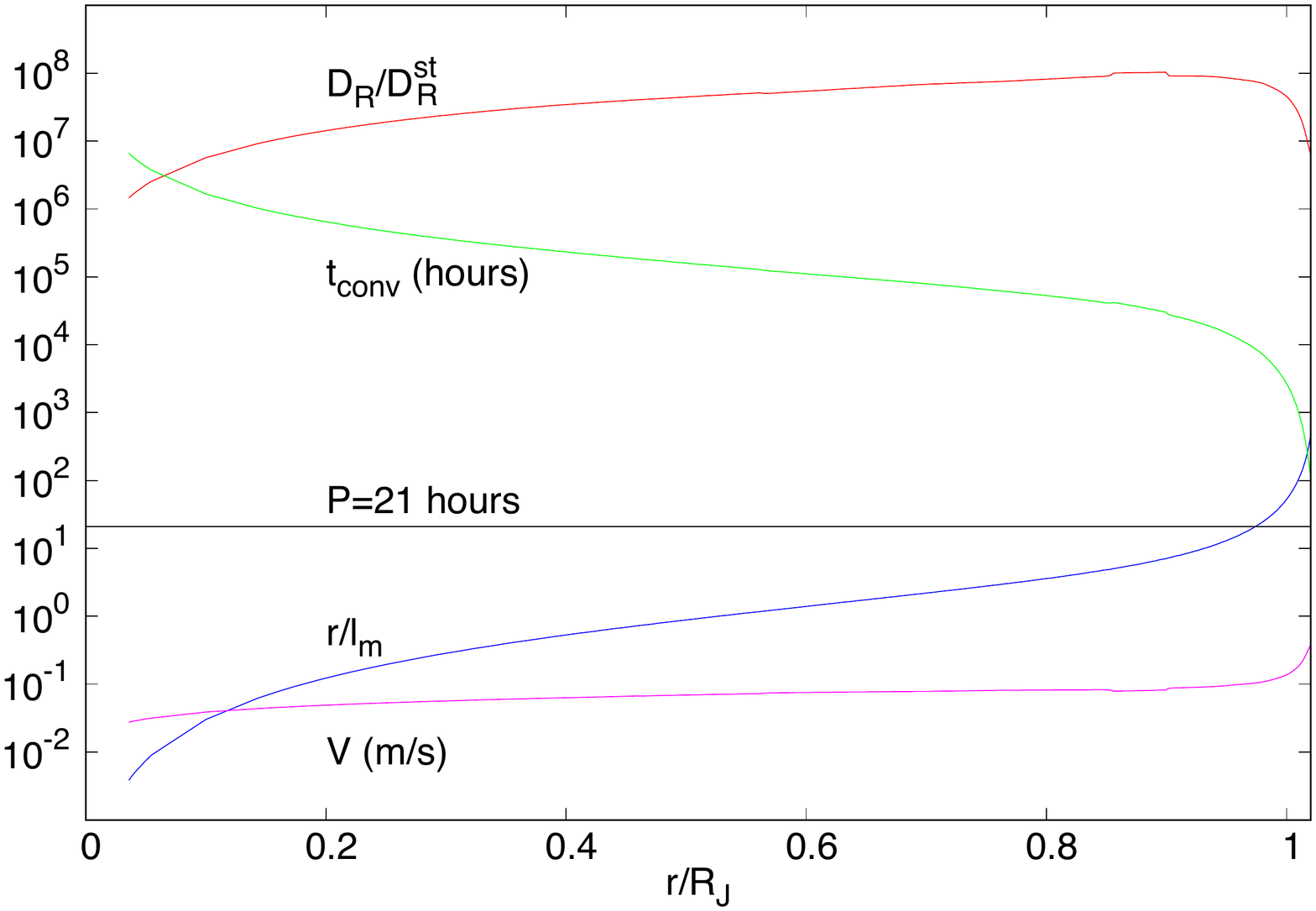}
    \caption{Atmosphere of Jupiter.  Shown are $r/l_m$ with $l_m=2H_P$ (blue curve),  the convective timescale $t_{\rm conv}$ in hours (green curve), the convective velocity $V$ in m~s$^{-1}$ (magenta curve) and $D_R/D^{\rm st}_R$ for $P=21$~hours (red curve) {\em versus} $r/R_{J}$, where $R_J$ is Jupiter radius,  using a vertical logarithmic scale and for a  model provided by I. Baraffe. The horizontal line  shows $P=21$~hours for comparison with $t_{\rm conv}$.  }
    \label{fig2_modeljupiter}
\end{figure*}

To calculate the rate of energy dissipation, we approximate the scale $H_c$ over which the convective velocity varies by the mixing length $l_m$, and use  the standard approximation $l_m = \alpha H_P$, where $\alpha =2$ and $H_P$ is the pressure scale height. 
Figure~\ref{fig2_modeljupiter} shows the convective timescale $t_{\rm conv}$, the convective velocity $V$, $r/l_m$  and $D_R/D^{\rm st}_R$ for $P=21$~hours  in the atmosphere of Jupiter, for a model provided by I. Baraffe (and described in \citealt{Baraffe2008}).     The model gives $H_P$ and the convective velocity $V$, calculated with the mixing length approximation, and we compute $t_{\rm conv} = l_m/V$.  This is not expected to be valid where $H_P >r$, which happens in the deep interior of Jupiter below $0.4 R_{\rm J}$, where $R_{\rm J}$ is Jupiter radius, as mixing length theory does not hold in this regime.  
However, as we will see below, the parts of the envelope below $0.4 R_{\rm J}$ do not contribute significantly to tidal dissipation.  

\begin{figure*}
    \centering
   \includegraphics[width=\columnwidth,angle=270]{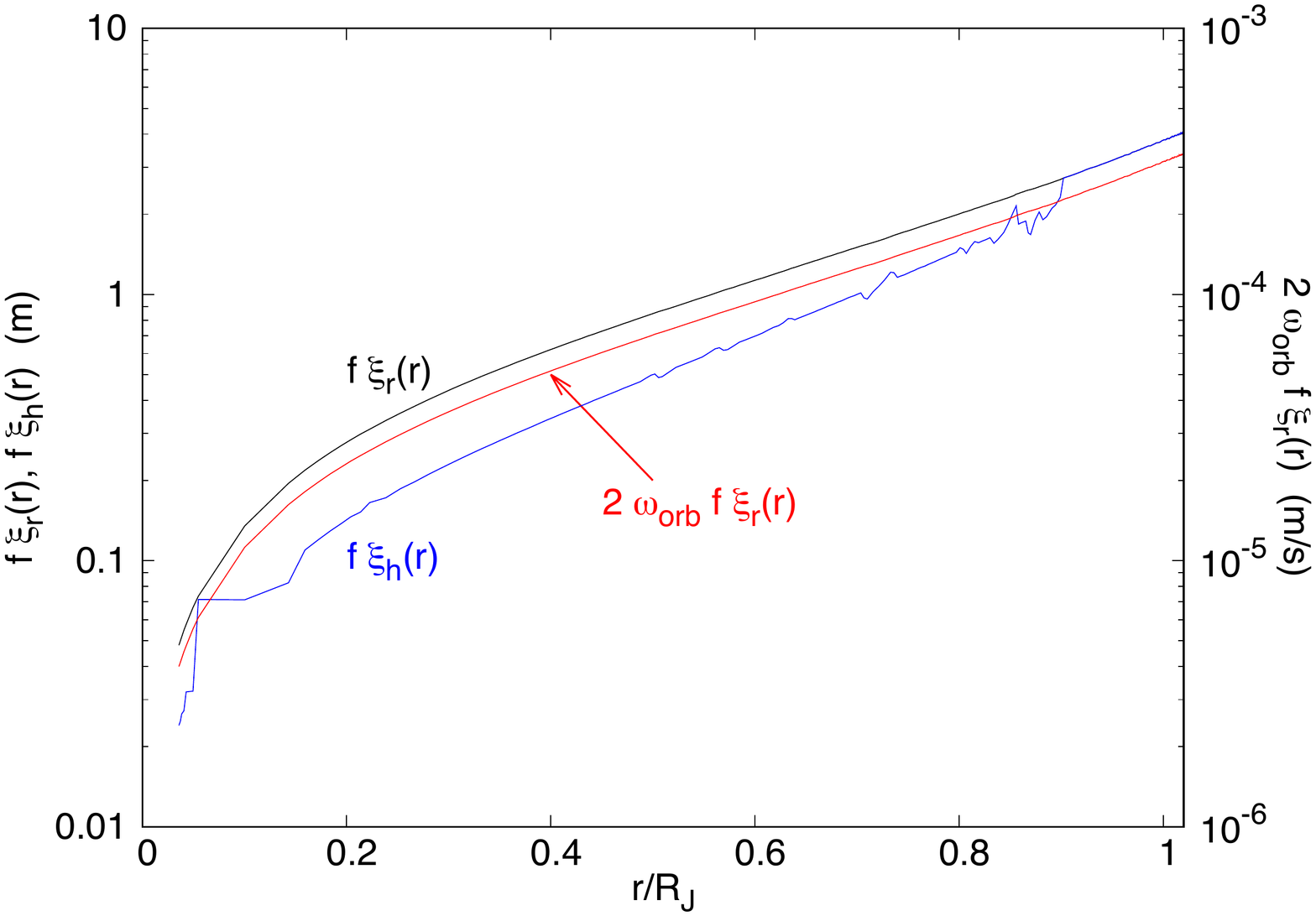}
    \caption{Equilibrium tide raised by Io in the atmosphere of Jupiter.  Shown  are $f \xi_r(r)$ and $f \xi_h(r)$ in m (black and blue curves, respectively, and left $y$--axis)  and  the radial part of $u'_r$, which is $2 \omega_{\rm orb} f \xi_(r)$, in m~s$^{-1}$ (red curve, right $y$--axis) {\em versus} $r/R_{J}$ using  vertical logarithmic scales.  As the data from Jupiter's model are noisy above 0.9$R_{J}$, we set $\xi_h(r) = \xi_r(r)$ there, which is a good approximation for the equilibrium tide. }
    \label{fig2a_tidejupiter}
\end{figure*}

Figure~\ref{fig2a_tidejupiter} shows  $f  \xi_r (r)$ and $f \xi_h (r)$, and the radial part of  $u'_r$, which is $ 2 \omega_{\rm orb}  f \xi_r(r)$, corresponding to the equilibrium tides given by equations~(\ref{eq:xirr}) and~(\ref{eq:xihr}), in the atmosphere of Jupiter.   As the data from Jupiter's model are noisy above 0.9$R_{J}$, we set $\xi_h(r) = \xi_r(r)$ there, which is a good approximation for the equilibrium tides when the interior mass is almost constant.

The effective tidal dissipation factor is defined as \citep{Goldreich1966}:
\begin{equation}
Q= \frac{2 \pi E} {\Delta E},
\label{eq:Qtidal}
\end{equation}
where $\Delta E$ is the energy lost by the tides during one tidal period, and $E$ is the energy stored in the tides themselves.  As there is equipartition between kinetic and potential energy, $E=2E_K$, where $E_K$ is the kinetic energy:
\begin{equation}
E_K = \iiint \frac{1}{2} \rho u'^2 r^2 \sin \theta {\rm d} r {\rm d} \theta {\rm d} \varphi,
\end{equation} 
where the integral is over the whole volume of Jupiter's atmosphere.  Using ${\bf u}' = \partial  {\bm \xi} / \partial t $, with ${\bm \xi}$ given by equations~(\ref{eq:xir})--(\ref{eq:xiphi}), yields:
\begin{equation}
E_K = \frac{24}{5} \pi  n^2 \omega^2 _{\rm orb} f^2  I_2 ,
\end{equation}
where:
\begin{equation}
I_2 = \int_{R_i}^{R_{\rm J}} \rho r^2 \left\{ \xi_r^2(r) + \frac{1}{6 r^2} \left[
\frac{{\rm d}}{{\rm d} r} \left( r^2 \xi_r(r) \right) \right]^2 \right\} {\rm d}r ,
\end{equation}
with $R_i$  being the inner radius of Jupiter's atmosphere.   

We now calculate $\Delta E =\left( {\rm d}E / {\rm d} t \right)_{\rm c} P$.  
We have argued in 
section~\ref{sec:dissipcircular} that,  when the body rotates rigidly,   ${\rm d}E/{\rm d}t$ is still given by equation~(\ref{eq:dEdt}),  but with the appropriate modification for the domain of integration of $I_1$.    As $t_{\rm conv}$ in Jupiter's atmosphere is everywhere much larger than the period of the tides relative to that of the fluid, $I_1$ is calculated by integrating over the whole atmosphere whether rotation is taken into account or not.  Therefore, rotation does not make a difference, and
 $\left( {\rm d}E / {\rm d} t \right)_{\rm c}$ is given by equation~(\ref{eq:dEdt}).  This yields:
\begin{equation}
Q=16 \omega_{\rm orb} \frac{I_2 }{ I_1},
\label{eq:QI1I2}
\end{equation}
where $I_1$ is given by equation~(\ref{eq:I1}).   Since $t_{\rm conv} \gg P$ everywhere in the atmosphere for all the periods  involving Jupiter's satellites, 
both $I_1$ and $I_2$ are independent of $\omega_{\rm orb}$, and $Q \propto \omega_{\rm orb}$.  For the orbital decay timescale,  equations~(\ref{eq:torb}) and~(\ref{eq:dEdtc}) yield $t_{\rm orb} \propto \omega_{\rm orb}^{-16/3}$.

For comparison, we see from equations~(\ref{eq:DRst}) and~(\ref{eq:convstress}) that standard mixing length theory gives ${\rm d}E / {\rm d}t \sim u'^2 \omega^{-s}_{\rm orb}$, where $s=1$ or 2 allows for suppression of dissipation at high frequency, and  $E=2E_K \sim u'^2$.  Therefore,  equation~(\ref{eq:Qtidal}) yields  
$Q \propto \omega^{s+1}_{\rm orb}$ when mixing length theory is used.  Note that, in this context,  a different scaling $Q \propto \omega^{s-1}_{\rm orb}$ was reported by \citet{Ogilvie2014}, based on the energy dissipation rate calculated by \citet{Zahn1977, Zahn1989}. The discrepancy arises from the fact that Zahn, following \citet{Darwin1879}, assumed that dissipation yielded a phase shift between the equilibrium tide and the tidal potential given by $\omega/ \left( t_{\rm conv} \omega^2_{\rm dyn} \right)$, where $\omega_{\rm dyn} = \left( GM/R^3 \right)^{1/2}$ is the dynamical frequency of the body in which the tides are raised, with  $M$ and $R$ being its mass and radius, respectively.  Such an assumption has not been used here, where we calculate the energy dissipation rate directly from equation~(\ref{eq:dEdttheta}) instead, replacing $D_R$ by $D^{\rm st}_R$ when using mixing length theory.

For the orbital frequency of Io, we obtain $E_K=2.0 \times 10^{27}$~ergs, $\left( {\rm d}E / {\rm d} t \right)_{\rm c} = 2.6 \times 10^{19}$~ergs~s$^{-1}$ and  $Q=  1.3 \times 10^4$.  This is  close to the value of $3.56 \times 10^4$ derived by \citet{Lainey2009} based on the orbital motion of the Galilean satellites.  As evidenced by the fact that Io is moving towards Jupiter, the orbital evolution of the Galilean satellites is dominated by the resonant interaction, and therefore the orbital evolution timescales cannot be calculated from equation~(\ref{eq:torb}).  

The upper part of Jupiter's atmosphere contributes significantly to $Q$: calculating $\left( {\rm d}E / {\rm d} t \right)_{\rm c}$ by including only the region below 0.9R$_{\rm J}$ yields $Q=5 \times 10^4$, whereas including only the region above 0.9R$_{\rm J}$  yields $Q=2.7 \times 10^4$.  This is because both $V/H_c = t^{-1}_{\rm conv}$ and the amplitude of the tides in equation~(\ref{eq:I1}) increase towards the surface.   The convective velocities at the surface of Jupiter may not be well approximated by the mixing length theory, but even if $V/H_c$ were smaller there we would still obtain $Q$ on the order of a few $10^4$.

We can write an approximate expression for $Q$ by noting that the tides enter the expressions for $\left( {\rm d}E / {\rm d} t \right)_{\rm c}$ and $E_K$ in a similar way.    Using $V/H_c = t^{-1}_{\rm conv}$ in equation~(\ref{eq:I1}), we can then 
approximate equation~(\ref{eq:Qtidal})  by:
\begin{equation}
Q \sim  n \omega_{\rm orb} \frac{ \int_{R_i}^{R_{\rm J}} \rho r^2  {\rm d} r }{ \int_{R_i}^{R_{\rm J}} t^{-1}_{\rm conv} \rho r^2  {\rm d} r}.
\label{eq:Qapprox}
\end{equation}
This yields $Q=  1.6 \times 10^4$, very close to the value obtained with equation~(\ref{eq:Qtidal}).  Although $\rho$ decreases towards the surface, $t_{\rm conv}$ decreases faster (while staying larger than the tidal period), so that the outer regions contribute most to $Q$.  The fact that $Q$ is well approximated by the expression above confirms that our results do not depend on the details of the components of the stress tensor we include in the calculation, as discussed in section~\ref{sec:timescales}.

\subsection{Saturn's tidal dissipation factor}

We now calculate the rate at which the energy of the tides raised by 
Enceladus in Saturn dissipate.  
This corresponds to $P_{\rm orb}=32.9$~hours and $M_p=1.08 \times 10^{20}$~kg.    Saturn's rotational period is 10.6~hours  but, as for Jupiter,  rotation can be neglected for calculating the tidal dissipation factor.  

As Saturn's models have been subject to recent developments, we use two different models, one provided by 
R.~Helled and A.~Vazan (model~1,  \citealt{Vazan2016}) and one provided by I.~Baraffe (model~2, \citealt{Baraffe2008}).

 Model~1 supplies the convective velocity $V_{\rm Sat}$, but it is not well resolved near the surface.  However, we find that $V_{\rm Sat}$ is  very close to $0.6V_{\rm Jup}$ in the bulk of the atmosphere, where $V_{\rm Jup}$ is the convective velocity output by the model of Jupiter  described above.  Therefore, for Saturn, we adopt the convective velocity $V=V_{\rm Sat}$ below $0.9R_{S}$ and $V=0.6V_{\rm Jup}$ above $0.9R_{S}$, where $R_{S}$ is Saturn radius.    As for Jupiter, we take the scale over which the convective velocity varies to be the mixing length $l_m= \alpha H_P$.  However, it has been argued that, in planetary interiors,  $\alpha$ may be smaller than the value of 2 commonly used in stellar physics (\citealt{Leconte2012}), and $V_{\rm Sat}$ in model~1 was calculated using  $\alpha=0.5$ \citep{Vazan2016}.  Therefore, we adopt $l_m=0.5H_P$.
 
Model~2 supplies $H_P$ and $V$, and we compute $t_{\rm conv} = l_m/V$ with $l_m=2H_P$, i.e. $\alpha=2$, as this is the value used to calculate $V$ in this model. 

\begin{figure*}
    \centering
   \includegraphics[width=\columnwidth,angle=270]{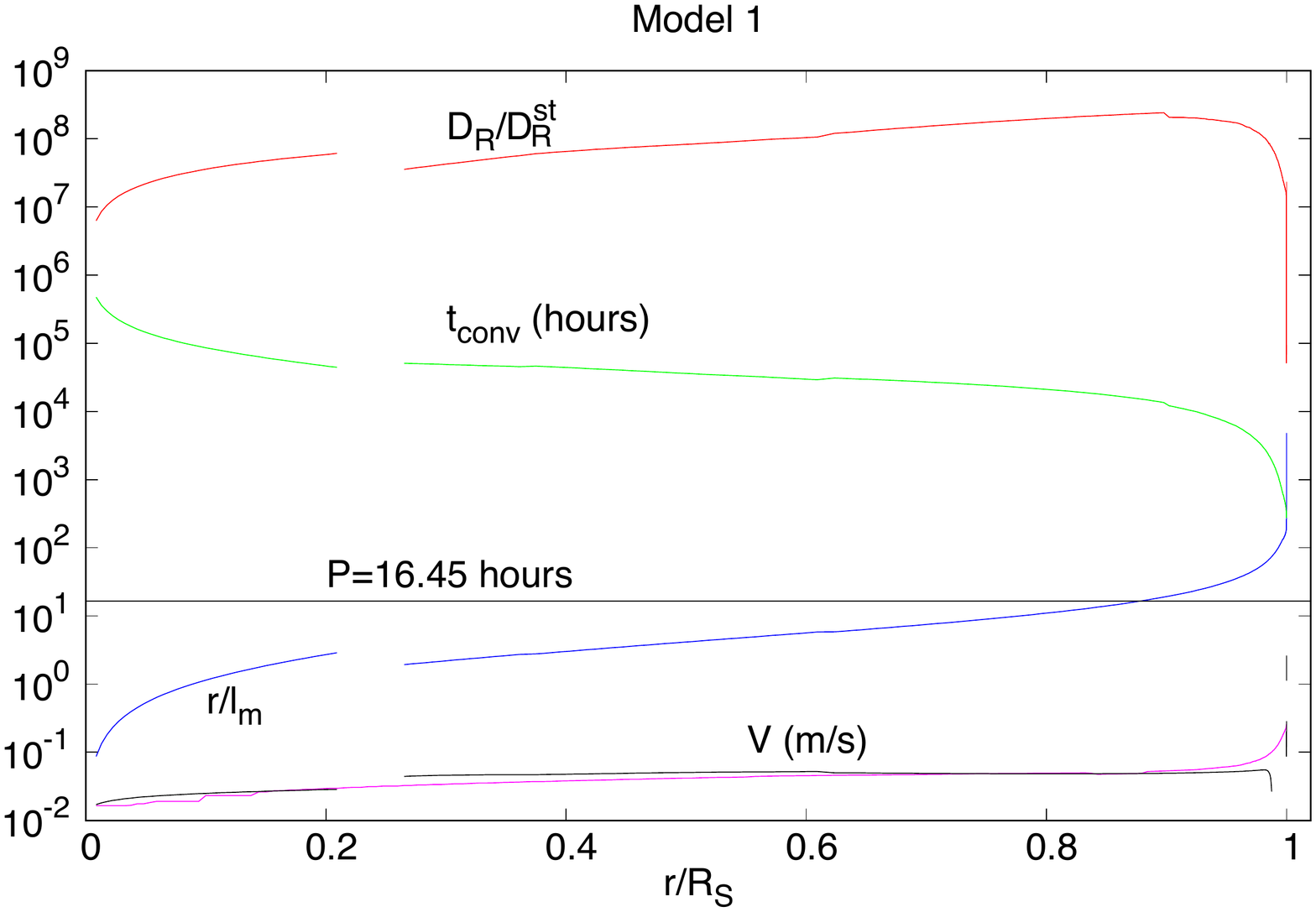}
    \includegraphics[width=\columnwidth,angle=270]{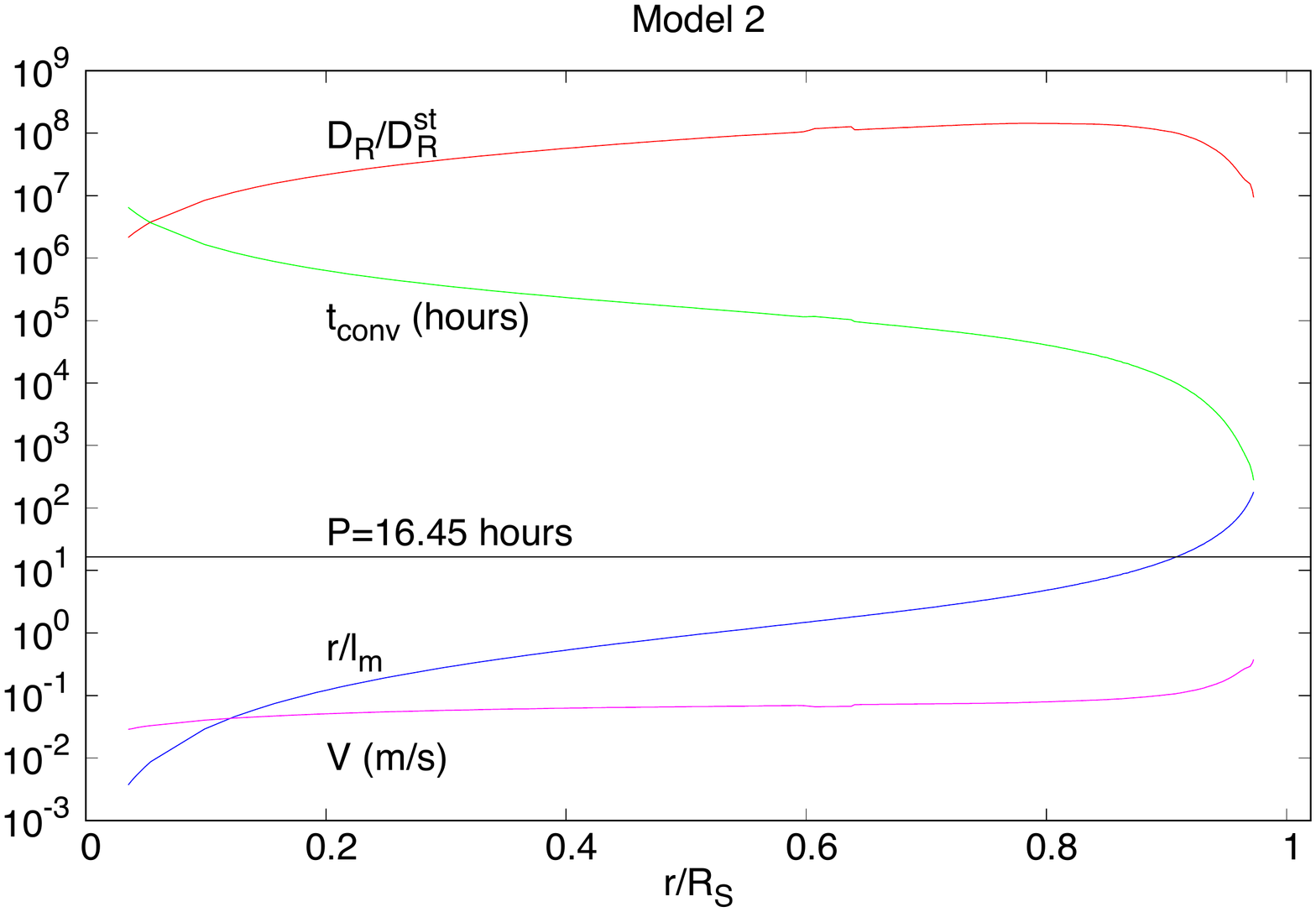}
    \caption{Atmosphere of Saturn.  {\em Model 1 (upper plot):} The black  and magenta curves show $V_{\rm Sat}$ and $0.6V_{\rm Jup}$, respectively,  in 
   m~s$^{-1}$,  {\em versus} $r/R_{S}$, where $R_S$ is Saturn radius.  We adopt $V=V_{\rm Sat}$ below $0.9R_{S}$ and $V=0.6V_{\rm Jup}$ above $0.9R_{S}$.  
Also
    shown are $r/l_m$ with $l_m=0.5H_P$ (blue curve),  the convective timescale $t_{\rm conv}$ in hours (green curve) and $D_R/D^{\rm st}_R$ for $P=16.45$~hours (red curve) {\em versus} $r/R_{S}$.  The vertical scale is logarithmic.  The model is provided by R.~Helled and A.~Vazan. The curves are interrupted in the regions which are stable against convection.   The horizontal line  shows $P=16.45$~hours for comparison with $t_{\rm conv}$.  {\em Model 2 (lower plot):}  Same as  upper plot but for a model provided by I.~Baraffe.  The magenta curve shows the convective velocity $V$ which is an output of the model and for this model $l_m=2H_P$. }
    \label{fig2_modelsaturn}
\end{figure*}

Figure~\ref{fig2_modelsaturn} shows $V_{\rm Sat}$, $0.6V_{\rm Jup}$,   $r/l_m$ with $l_m=0.5H_P$, $t_{\rm conv}=l_m/V$  and $D_R/D^{\rm st}_R$ for $P=16.45$~hours  for model~1, and $V$,   $r/l_m$ with $l_m=2H_P$, $t_{\rm conv}=l_m/V$  and $D_R/D^{\rm st}_R$ for model~2.    Note that  model~1 has regions which are stable against convection (\citealt{Leconte2012, Leconte2013}, \citealt{Vazan2016}).


Using astrometric observations spanning more than a century together with Cassini data,  
\citet{Lainey2017} have recently determined the effective tidal dissipation factor $Q$ for Saturn interacting with its moons Enceladus, Tethys, Dione and Rhea, which have orbital periods of 1.37, 1.89, 2.74 and 4.52~days, respectively.    Using equation~(\ref{eq:Qtidal}) and model~1, we find $Q_{\rm Encel}=4.5 \times 10^3$  for Saturn interacting with Enceladus.   This  is in very  good agreement with the value  published by \citet{Lainey2017}, which is $2.45 \times 10^3$.   As Enceladus is closer to Saturn than Dione, and $t_{\rm orb} \propto \omega_{\rm orb}^{-16/3}$,  its interaction with Saturn yields a shorter orbital decay timescale than that for Dione.  However, the two moons are dynamically coupled through a 2:1 mean motion  resonant interaction, which implies that they both migrate at the same rate corresponding to the strongest interaction with Saturn.  Therefore, $ Q_{\rm Dione} \sim Q_{\rm Encel} $, consistent with \citet{Lainey2017}.  Although these authors do not measure an orbital evolution timescale for Mimas, this moon is in a 4:2 mean motion resonance with Tethys, so the $Q$ value for both satellites interacting with Saturn should be the same, equal to that of Mimas.  Using equation~(\ref{eq:Qtidal}), we obtain $Q_{\rm Mimas}=6.5 \times 10^3$, which is 1.4 times larger than $Q_{\rm Encel}$,  in excellent agreement with the ratio $Q_{\rm Tethys}/Q_{\rm Encel}=1.3$ reported by \citet{Lainey2017}.   For Rhea, we obtain $Q_{\rm Rhea} =1.4 \times 10^3$, which is about 4 times larger than the value of 315 reported by \citet{Lainey2017}.  Note that, as for Jupiter, $Q \propto \omega_{\rm orb}$. 

Model 2 with $\alpha=2$ yields $Q_{\rm Encel}=1.6 \times 10^4$.  I.~Baraffe also provided model~2 with convective velocities calculated adopting $\alpha=0.5$. Using $l_m=0.5H_P$ with this model yields  $Q_{\rm Encel}=7.3 \times 10^3$.    In addition to model~1, R.~Helled  supplied several models which were calculated with a planetary evolution code, as described in \citet{Vazan2016}.  Finally, Y.~Miguel and T.~Guillot  provided a model which matches all the gravity harmonics measured by Cassini,  mass, radius and differential rotation \citep{Galanti2019}.  These models do not output the convective velocities, so we used $V=0.6V_{\rm Jup}$.  The values of $Q$ obtained in all cases were consistent with the results described above.  Therefore, tidal dissipation in Saturn is not sensitive to the details of the structure, but to the values of the convective timescale.
This is consistent with the fact that $Q$ is well approximated by equation~(\ref{eq:Qapprox}).

This suggests that, if tidal dissipation of the equilibrium tides is responsible for the orbital evolution of Saturn's moons, the mixing length parameter in Saturn's interior may be smaller  than the commonly assumed value of $\alpha=2$, in agreement with the models of \citet{Vazan2016}.

\subsection{Circularization of late--type binaries}
\label{sec:sun}

In the literature, the  effective tidal dissipation factor has been used for stars as well as for giant planets.  However, it is not an easy quantity to calculate for stars, because the energy stored in the tides cannot be evaluated using the equilibrium approximation (see, e.g., \citealt{Terquem1998}).   Also, since the tides are dissipated in only part of the star, while the energy $E_K$ requires integration over the entire volume of the star, $Q$ depends on the amplitude of the tides and therefore has a less straightforward dependence on $\omega_{\rm orb}$ than in giant planets.   For this reason, we will not compute values of $Q$ in this section.  

\begin{figure*}
    \centering
   \includegraphics[width=\columnwidth,angle=270]{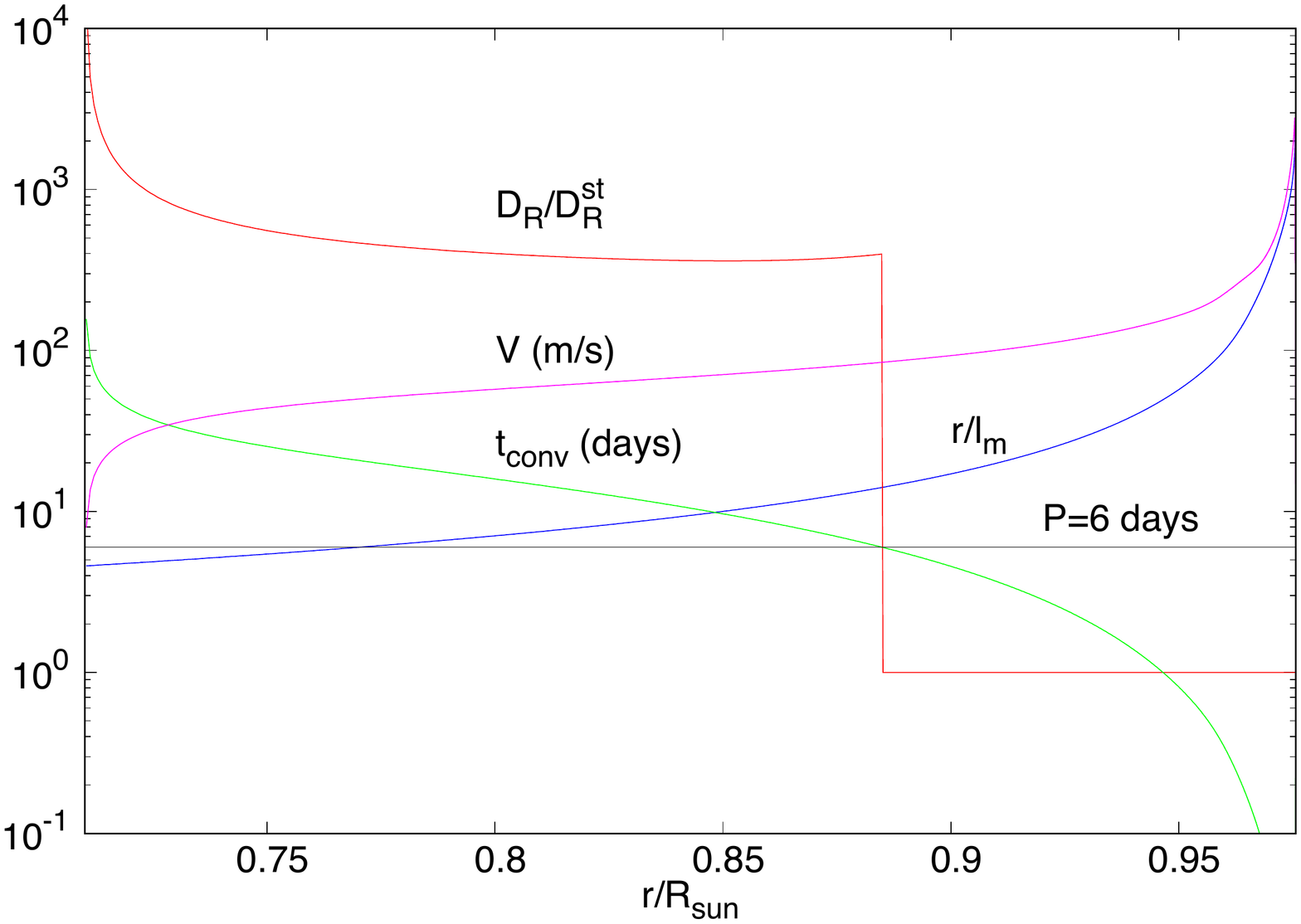}
    \caption{Convective envelope of the Sun.  Shown are $r/l_m$ with $l_m=2H_P$ (blue curve),  the convective timescale $t_{\rm conv}$ in days (green curve), the convective velocity $V$ in m~s$^{-1}$ (magenta curve) and $D_R/D^{\rm st}_R$  for $P=6$~days  (red curve) {\em versus} $r/R_{\sun}$ using a vertical logarithmic scale and for a 1~M$_{\sun}$ MESA model.  The horizontal line  shows $P=6$~days for comparison with $t_{\rm conv}$, and we set  $D_R/D^{\rm st}_R=1$ where  $t_{\rm conv} <P$. }
    \label{fig1_modelsun}
\end{figure*}

The results presented in this section have been obtained using a solar
 model produced by MESA \citep{Paxton2011, Paxton2013,Paxton2015, Paxton2016, Paxton2018, Paxton2019}, and have been checked not to differ from those obtained using a solar model provided by I. Baraffe.   The code outputs  the pressure scale height $H_P$ and the convective velocity $V$ computed with the mixing length theory and using $\alpha=2$.  We note $l_m=2H_P$ the mixing length.  
Figure~\ref{fig1_modelsun} shows the convective timescale $t_{\rm conv}=l_m/V$, the convective velocity $V$, $r/l_m$  and $D_R/D^{\rm st}_R$ for $P=6$~days ($P_{\rm orb}=12$~days) and using $H_c=l_m$ in the convective zone.   

\begin{figure*}
    \centering
   \includegraphics[width=\columnwidth,angle=270]{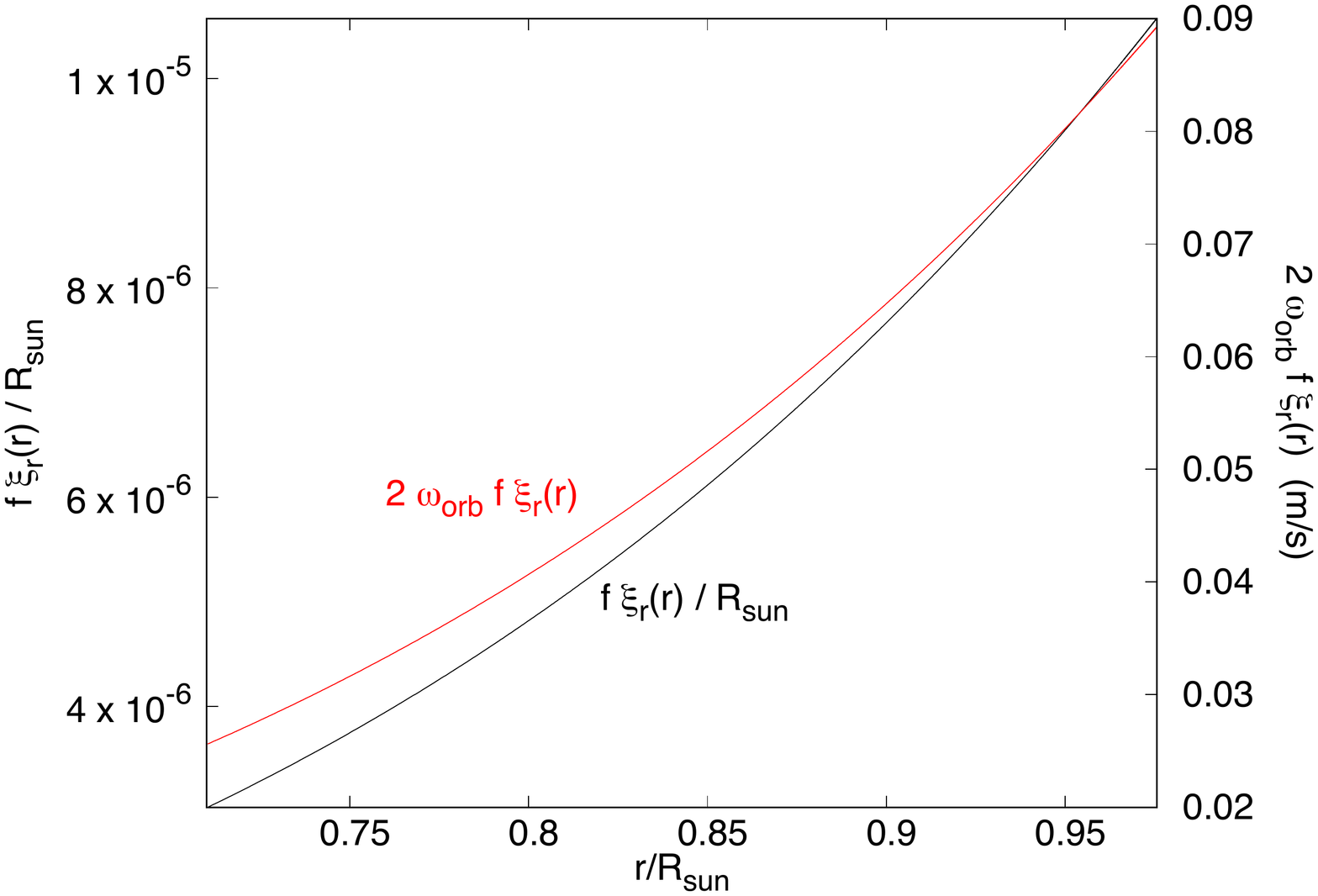}
    \caption{Equilibrium tide raised in the convective envelope of the Sun by a 1~M$_{\sun}$ mass star for an orbital period of 12 days.  Shown  are $f \xi_r(r) / R_{\sun}$  in m (black  curve, left $y$--axis)  and  the radial part of $u'_r$, which is $2 \omega_{\rm orb} f \xi_r(r)$, in m~s$^{-1}$ (red curve, right $y$--axis) {\em versus} $r/R_{\sun}$.  In the convective envelope of the Sun,  $\xi_h(r) \simeq \xi_r(r)$. }
    \label{fig1_tidesun}
\end{figure*}

Figure~\ref{fig1_tidesun} shows  $f  \xi_r (r) / R_{\sun}$  and the radial part of  $u'_r$, which is $ 2 \omega_{\rm orb}  f \xi_r(r)$, corresponding to the equilibrium tides given by equations~(\ref{eq:xirr}) and~(\ref{eq:xihr}), in the convective envelope of the Sun.   As the mass interior to  radius $r$ varies slowly  with $r$, $\xi_h(r) \simeq \xi_r(r)$ there.  

\begin{figure*}
    \centering
   \includegraphics[width=\columnwidth,angle=270]{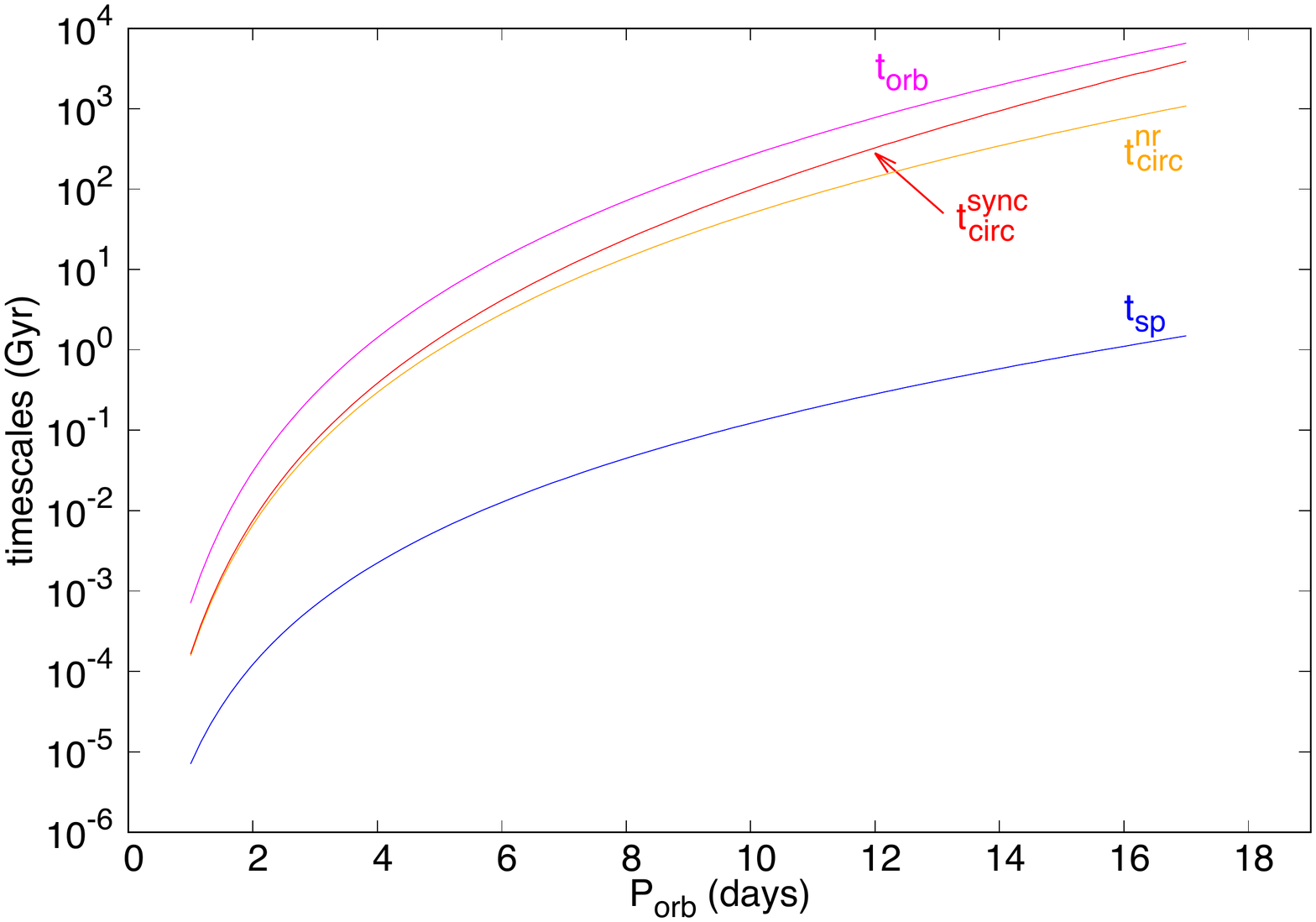}
    \caption{Tides raised in a 1~M$_{\sun}$ MESA model  by a companion with $M_p =1$~M$_{\sun}$.
    Shown are, from top to bottom,  the orbital decay timescale (magenta curve),   circularization timescale for a synchronized star (red curve),  circularization timescale for a non--rotating star (orange curve) and spin up timescale (blue curve)  in Gyr, using a logarithmic scale,  {\em versus} orbital period in days.   The curves correspond  to 
    the timescales calculated from equations~(\ref{eq:torb}), (\ref{eq:tcircsync}), (\ref{eq:tcircnr}), and~(\ref{eq:tsp}), respectively.  The circularization timescales calculated that way are about 40 times too large to account for the circularization of late--type binaries.  
      }
    \label{fig4}
\end{figure*}

For the 1~M$_{\sun}$ MESA model represented in figure~\ref{fig1_modelsun},  writing the moment of inertia as $I=0.07 M_{\sun} R^2_{\sun}$ and using $H_c=l_m$, we calculate the timescales 
 given by equations~(\ref{eq:torb}), (\ref{eq:tsp}), (\ref{eq:tcircnr})  and~(\ref{eq:tcircsync}) and display them in figure~\ref{fig4}.   The circularization timescales calculated that way are about 40 times too large to account for the circularization of late--type binaries. 

As seen from equation~(\ref{eq:DRtide}), the circularization timescale we obtain here is about 
$\left( r/H_c \right)^2$ larger than the timescale $t^{\rm st}_{\rm circ}$ obtained with the standard approach when suppression of dissipation by large eddies is ignored.  However,  $t^{\rm st}_{\rm circ}$ is orders of magnitude too large to account for the circularization timescale of late--type binaries (\citealt{Goodman1997}, \citealt{Terquem1998}), and as $r/H_c $ is  only between about 5 and 10 in the region of the convective envelope where $t_{\rm conv} >P$, if we use $H_c=l_m$, the timescale using the new formalism is still too long.  

It is not clear how the timescales could be decreased by a factor of 40  within the context of the mechanism discussed here. Only by replacing the shear rates $V/H_c$ and $V/r$ by  $V/P$ in equation~(\ref{eq:I1}) and integrating over the whole extent of the convective zone do we get timescales matching observations.   
Therefore, circularization of late--type binaries may occur as a result of other processes  than the interaction between convection and equilibrium tides.   The strong shear at the bottom of the convective zone, where the convective velocity rapidly reaches zero, or in the tachocline, where the rotational velocity has a strong radial gradient, may contribute to the dissipation of tides.  

\begin{figure*}
    \centering
   \includegraphics[width=\columnwidth,angle=270]{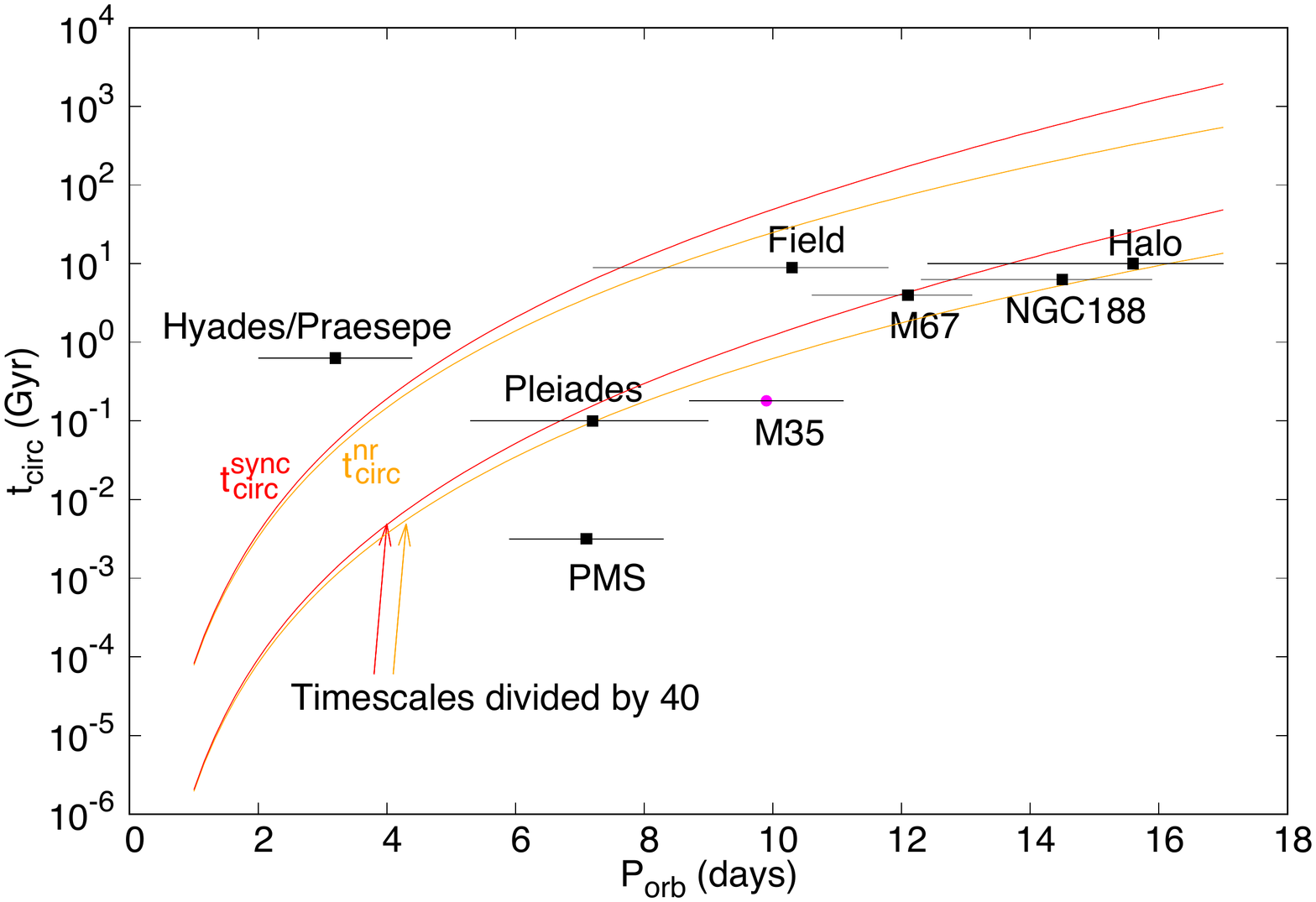}
    \caption{Late--type binaries.  Shown are the circularization timescales in Gyr and the same timescales divided by 40 using a logarithmic scale for non--rotating stars (orange curves) and synchronized stars (red curves) {\em versus} orbital period in days.   The timescales are calculated from equations~(\ref{eq:tcircnr}) and~(\ref{eq:tcircsync}) using a 1~M$_{\sun}$ MESA model, and the results are divided by two  assuming the binary is made of two identical stars.   The black crosses with error bars represent data from \citet{Meibom2005}, whereas the magenta filled circle is from  \citet{Leiner2015}.
      }
    \label{fig3}
\end{figure*}

As the formalism presented above yields a $Q$ factor for Jupiter and possibly for Saturn in good agreement with observations, it may apply to the interior of giant planets.  To be able to infer the orbital evolution of binaries containing a star and a hot Jupiter, we therefore scale the timescales  resulting from the tides raised in the star so that they match the observations for late--type binaries.  This is shown in figure~\ref{fig3}, where we plot the timescales given by equations~(\ref{eq:tcircnr}) and~ (\ref{eq:tcircsync}), using a 1~M$_{\sun}$ MESA model and $M_p=1$~M$_{\sun}$, divided by 40, together with  data showing the  circularization period {\em versus} age for eight late--type binary populations (note that the timescales are divided by 2 before applying the scaling as the binary is assumed to have two identical stars).  

We display  the circularization timescale for both non--rotating and synchronized stars.  However, 
 from figure~\ref{fig4}, we expect  the stars to be synchronized on a relatively short timescale, so that when comparing with observations the timescale for synchronized stars should be used. 
Note that solar type stars on the main sequence lose angular momentum because of magnetized winds.  \citet{Gallet2013} derive a corresponding timescale $J/ \left| {\rm d}J / {\rm d}t \right|$, where $J$ is the stellar angular momentum,  on the order of a few Gyr for stars which are a few Gyr old.  As this is much longer than the tidal spin up timescale (especially after the scaling is applied), we would expect  tidal synchronization to be achieved despite braking of the stars by winds.

All the data except that for M35  are taken from \citet{Meibom2005}.      For M35, the circularization period of 9.9~days is from \citet{Leiner2015} and the age of 0.18~Gyr from \citet{Kalirai2003}.
For
PMS binaries, our calculation does not actually apply,  because those stars have a more extended convective envelope than the Sun.  Tides are therefore more efficiently dissipated in those stars, leading to shorter circularization timescales for a given period.  A proper calculation for PMS binaries is done in section~\ref{sec:PMS}.  For  Hyades/Praesepe, the circularization period makes this cluster very unusual, but it is worth noting that it is based on a small sample.  
For field binaries, there is also some discrepancy between the results presented here after scaling and the data published by \citet{Meibom2005}.  However, these authors point out that the age of this population is not well constrained, which makes the sample not very reliable.  Also, a survey published by \citet{Raghavan2010} report a circularization period close to 12~days, which would move the data point for this population closer to the curves in figure~\ref{fig3}. 

The timescales 
 given by equations~(\ref{eq:torb}), (\ref{eq:tsp}), (\ref{eq:tcircnr})  and~(\ref{eq:tcircsync}) and divided by 40
can be fitted by the following power laws for  $1 \le P_{\rm orb} \le 17$~days:
\begin{align}
t_{\rm orb} ({\rm Gyr}) & =  2.175  \; \frac{\left( 1+M_p/M_c \right)^{5/3}}{M_p/M_c}  \left( \frac{P_{\rm orb}}{10~{\rm days}} \right)^{5.695}  , 
\label{eq:fittorb} \\
t_{\rm sp} ({\rm Gyr}) & =  7.997 \times 10^{-4}  \left( 1+\frac{M_c}{M_p} \right)^{2} \left( \frac{P_{\rm orb}}{10~{\rm days}} \right)^{4.362}  , \\
t^{\rm nr}_{\rm circ} ({\rm Gyr}) & =  0.403 \; \frac{\left( 1+M_p/M_c \right)^{5/3}}{M_p/M_c} \left( \frac{P_{\rm orb}}{10~{\rm days}} \right)^{5.586}  , \\
t^{\rm sync}_{\rm circ} ({\rm Gyr}) & =  0.867  \; \frac{\left( 1+M_p/M_c \right)^{5/3}}{M_p/M_c}  \left( \frac{P_{\rm orb}}{10~{\rm days}} \right)^{6.054}  , \label{eq:fittcircsync}
\end{align}
where the dependence on $M_p$ is shown explicitly.  The ratio of these fits to the original timescales is between 0.5 and 1.3.  

In calculating ${\rm d}E / {\rm d}t$, we have neglected the contribution from $D^{\rm st}_R$ in the region where $t_{\rm conv} < P$.  This becomes important when the timescale $t^{\rm st}_{\rm circ}$ obtained with the standard approach and ignoring suppression of dissipation by large eddies becomes comparable to the circularization timescale we calculate here.  We have checked that this is the case only for the largest orbital period of 17~days considered here. 

\subsection{Circularization of pre--main sequence binaries}
\label{sec:PMS}

\begin{figure*}
    \centering
   \includegraphics[width=\columnwidth,angle=270]{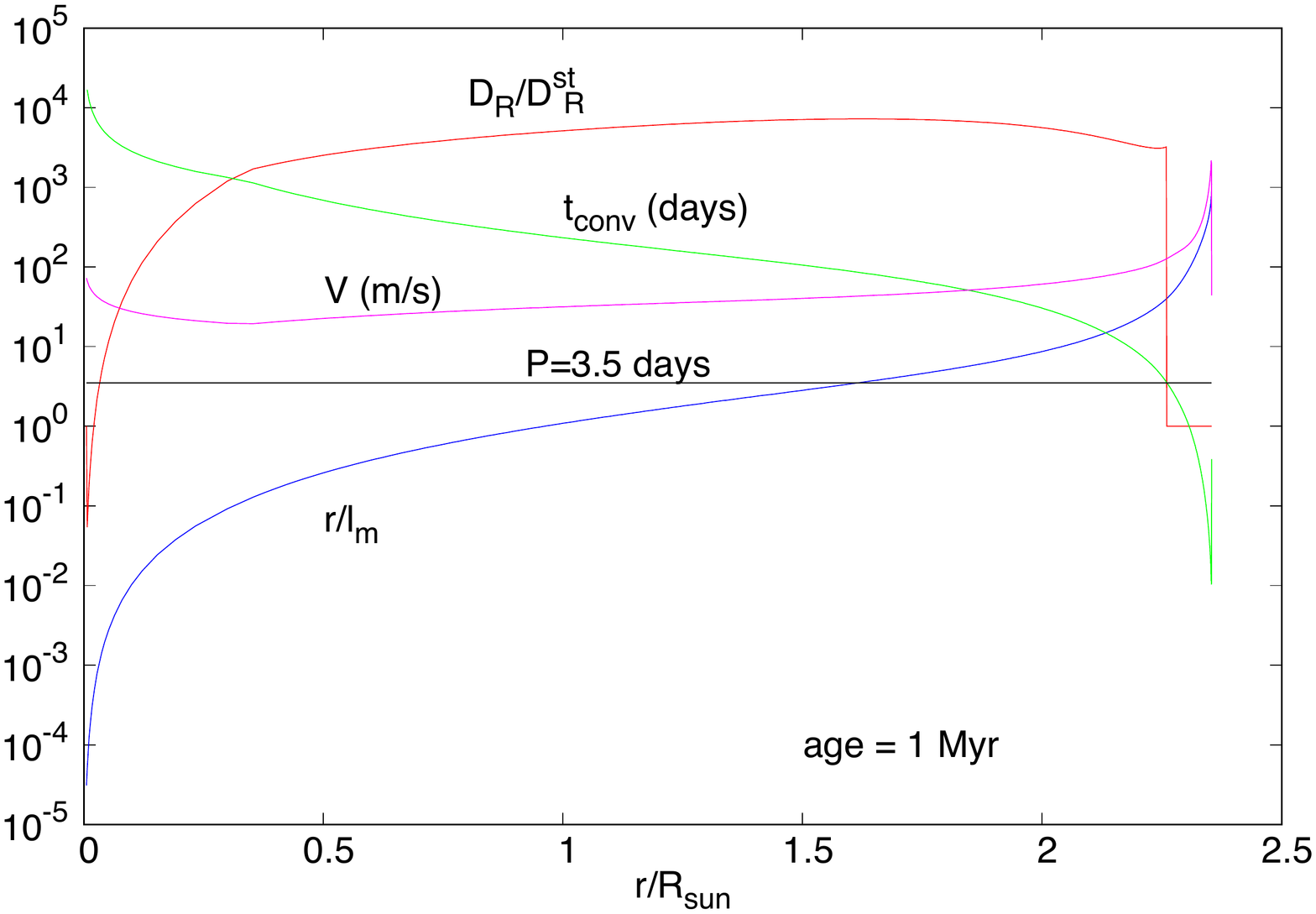}
    \caption{1~Myr old PMS star.  Shown are $r/l_m$ with $l_m=2H_P$ (blue curve),  the convective timescale $t_{\rm conv}$ in days (green curve), the convective velocity $V$ in m~s$^{-1}$ (magenta curve) and $D_R/D^{\rm st}_R$  for $P=3.5$~days  (red curve) {\em versus} $r/R_{\sun}$ using a vertical logarithmic scale and for a 1~M$_{\sun}$ MESA model.  The horizontal line  shows $P=3.5$~days for comparison with $t_{\rm conv}$, and we set  $D_R/D^{\rm st}_R=1$ where  $t_{\rm conv} <P$.  The star is completely convective.}
    \label{fig1_modelPMS}
\end{figure*}

We generate models of 1~M$_{\sun}$ PMS stars of different ages using MESA.  
Figure~\ref{fig1_modelPMS} shows the convective timescale $t_{\rm conv}=l_m/V$, the convective velocity $V$, $r/l_m$  and $D_R/D^{\rm st}_R$ for $P=3.5$~days ($P_{\rm orb}=7$~days) and using $H_c=l_m=2H_P$ for a 1~Myr old star.  The star has a radius of 2.35~R$_{\sun}$ and is completely convective.  For a 3.16~Myr old star, the radius   is 1.63~R$_{\sun}$ and the convective envelope only extends down to about 0.3 stellar radius.  

Assuming a binary with two identical stars, we calculate the orbital period $P_{\rm circ}$ for which the circularization timescale is equal to the age $t_{\rm age}$ of the stars.  For PMS binaries, the timescales corresponding to non--rotating and synchronous stars are roughly the same, so they are calculated from either equation~(\ref{eq:tcircnr}) or~(\ref{eq:tcircsync}) and divided by  two to account for the two stars.    We find $P_{\rm circ}=5.2$, 5.9 and 7.3~days for $ t_{\rm age}=3.16$, 2 and 1~Myr, respectively.  Younger stars  would give larger $P_{\rm circ}$, but our calculations are probably not valid when a massive disc is still present around the stars, which is the case during the first Myr or so.    Therefore, our results indicate that binaries circularize early  on during the PMS phase up to a period of about 7 days, which is in good agreement with the observed period of 7.1~days for the PMS population shown in figure~\ref{fig3} and which has an age of 3.16~Myr.

\subsection{Hot Jupiters}

\begin{figure*}
    \centering
   \includegraphics[width=\columnwidth,angle=270]{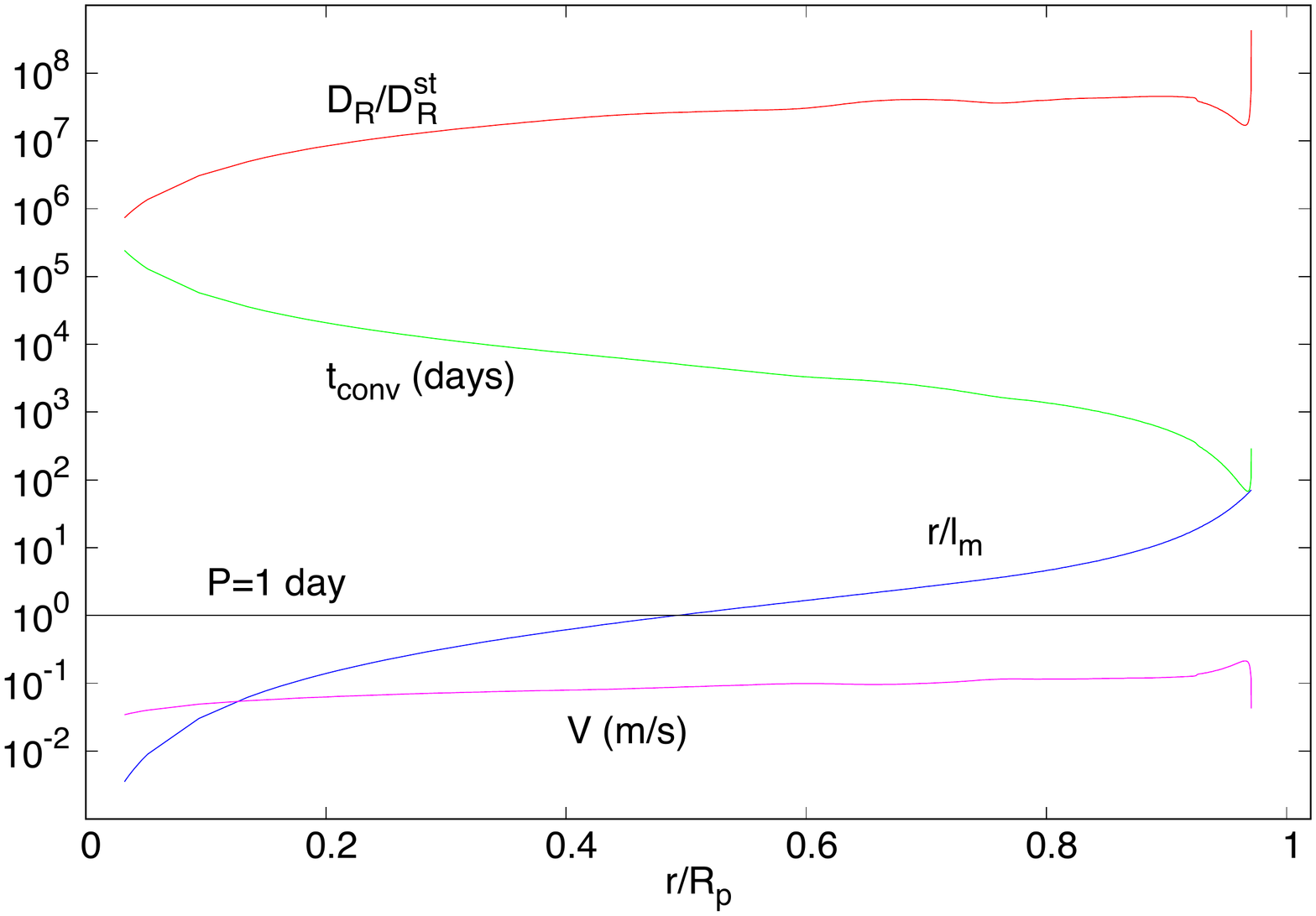}
    \caption{Atmosphere of an irradiated Jupiter.  Shown are $r/l_m$ with $l_m=2H_P$ (blue curve),  the convective timescale $t_{\rm conv}$ in days (green curve),  the convective velocity $V$ in m~s$^{-1}$ (magenta curve)  and $D_R/D^{\rm st}_R$ for $P=1$~day (red curve) {\em versus} $r/R_{\rm p}$, where $R_{\rm p}=1.126$~R$_{\rm J}$ is the planet  radius,  using a vertical logarithmic scale. The horizontal line  shows $P=1$~day for comparison with $t_{\rm conv}$.  }
    \label{fig6}
\end{figure*}

We now consider the case where the central mass is a solar type star and the companion a Jupiter mass planet.   As we are interested in planets which are close to their host star, we use a model for an irradiated Jupiter.
Figure~\ref{fig6} shows the convective timescale $t_{\rm conv}$, the convective velocity $V$, $r/l_m$ with $l_m=2 H_P$ and $D_R/D^{\rm st}_R$ for $P=1$~day ($P_{\rm orb}=2$~days)  in the atmosphere of an irradiated Jupiter, for a model  provided by I.~Baraffe.  This model corresponds to a planet which has an orbital period of about 2~days  around an F~star, which is slightly hotter than the Sun.  It has a (non--inflated) radius  $R_{\rm p}=1.126$~R$_{\rm J}$, and there is a radiative layer near the surface due to irradiation.  This model is more irradiated than the planets which would be consistent with the parameters we adopt here.  However, by calculating results for both this model and a standard Jupiter, we can bracket all realistic models.  For the moment of inertia of the planet, we adopt $I=0.27M_{\rm J} R^2_{\rm p}$, which gives Jupiter's value when $R_{\rm p}=R_{\rm J}$.

\begin{figure*}
    \centering
   \includegraphics[width=\columnwidth,angle=270]{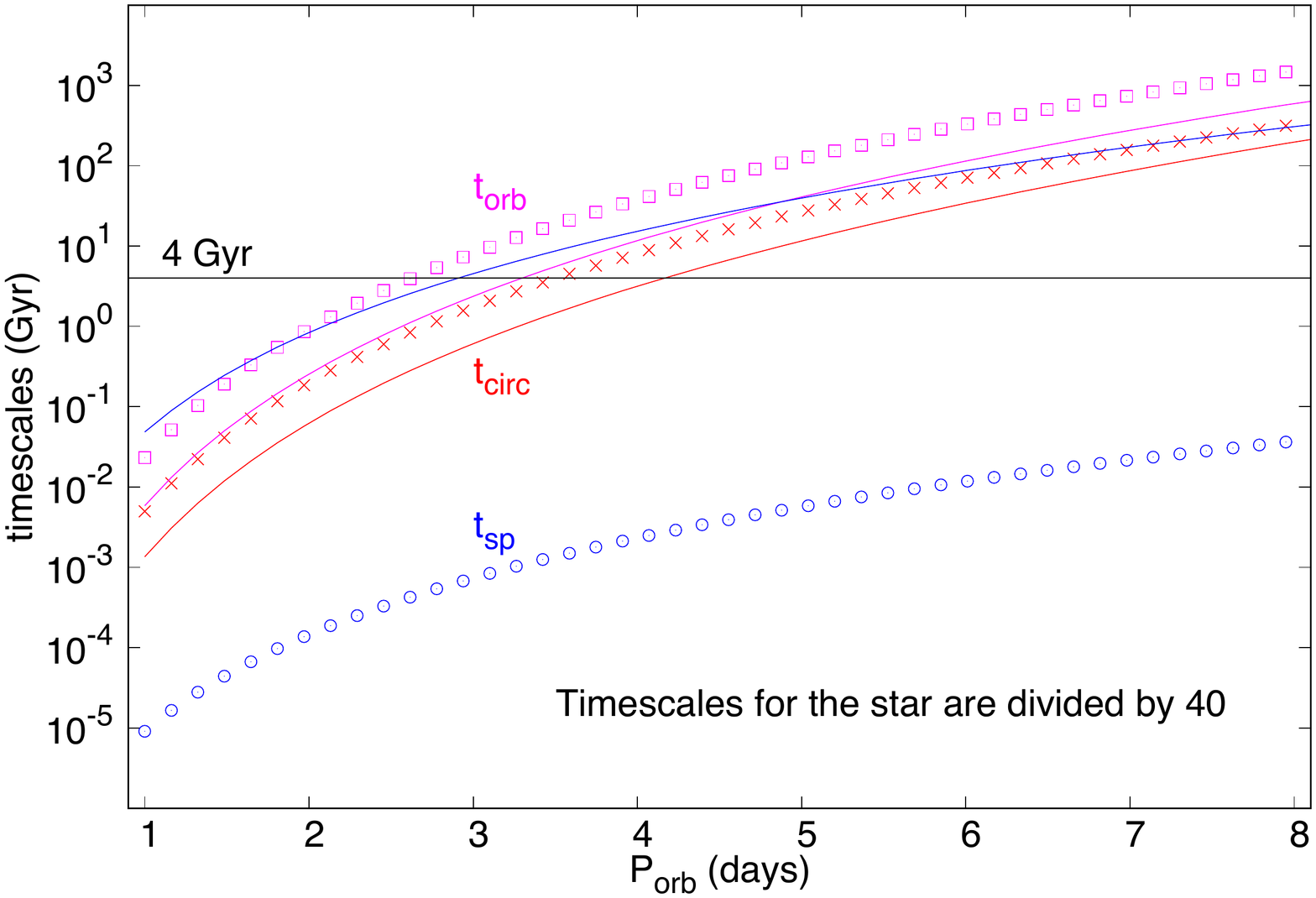}
    \caption{Hot Jupiters.  Shown are the orbital decay (magenta),    circularization   (red)  and spin up  (blue) timescales  in Gyr using a logarithmic scale  {\em versus} orbital period in days.  The solid curves correspond to the tides raised in a 1~M$_{\sun}$ star by a Jupiter mass planet, and the symbols correspond to the tides raised in an irradiated Jupiter mass planet by a  1~M$_{\sun}$ star.   
    The timescales are
 given by equations~(\ref{eq:torb}), (\ref{eq:tsp}) and (\ref{eq:tcircnr})  (or equivalently eq.~[\ref{eq:tcircsync}], as  $t^{\rm sync}_{\rm circ} \simeq t^{\rm nr}_{\rm circ}$ for both the tides raised in the star and the planet), and have been divided  by 40 for the tides raised in the star.   
        The line corresponding to a timescale of 4~Gyr is also shown  to indicate the periods at which circularization and synchronization occur on such a timescale.
      }
    \label{fig5}
\end{figure*}

Figure~\ref{fig5}  shows  the circularization, orbital decay and spin up  timescales  {\em versus} orbital period between 1 and 8 days corresponding to both the tides raised in the star by the planet and the tides raised in the planet by the star.   
The timescales are
 given by equations~(\ref{eq:torb}), (\ref{eq:tsp}), (\ref{eq:tcircnr})  and~(\ref{eq:tcircsync}), and have been divided  by 40 for the tides raised in the star.  At the short periods of interest here,  $t^{\rm sync}_{\rm circ} \simeq t^{\rm nr}_{\rm circ}$ for both the tides raised in the star and the planet,  as $t_{\rm conv}$ is large enough compared  to the tidal period that the dominant term $I_1 \left( \omega_{\rm orb}, 2, 3 \right) \simeq  I_1 \left( \omega_{\rm orb},2,1 \right) $ in equation~(\ref{eq:tcircnr}).     


Circularization and orbital decay occur predominantly as a result of the tides raised in the star, and the tides raised in the planet are only important to synchronize it.   We have checked that replacing the irradiated Jupiter model by the standard Jupiter model described above led to very similar results, with timescales corresponding to the tides raised in the planet being 1.3 to 1.7 times longer.  

{\em Circularization:}  Figure~\ref{fig5} shows that  the orbit of hot Jupiters should circularize up to periods of 4--5~days on timescales of a few Gyr.    These results are in  agreement with observations, which indicate a circularization period of  5--6~days \citep{Halbwachs2005, Pont2009, Pont2011}.    

{\em Synchronization:}  Due to the tides raised by the star, the planet is synchronized on timescales much shorter than the age of the systems.   Our results indicate that, for periods below about 3~days, the star itself should synchronize on timescales of at most a few Gyr because of the tides raised by the planet.  However, as already pointed out above, solar type stars on the main sequence lose angular momentum because of magnetized winds.  The corresponding timescale $J/ \left| {\rm d}J / {\rm d}t \right|$, where $J$ is the stellar angular momentum,  is on the order of a few Gyr for stars which are a few Gyr old, and much smaller for younger stars \citep{Gallet2013}.  This is shorter or equal to the spin up timescales found here.   Therefore, this braking of the star by winds may prevent tidal synchronization by hot Jupiters.  This is suggested by observations which show that, although stars hosting hot Jupiters spin faster than similar  stars without companions, they are not synchronized 
\citep{Penev2018}.

{\em Orbital decay:}  From figure~\ref{fig5}, we see that orbital decay becomes significant for periods below  3--4~days.    If both the star and the planet were synchronized and the orbit circular, orbital evolution would not occur.  However, as pointed out above, stars with hot Jupiters are not observed to be synchronized, so that our results imply that orbital decay  occurs in these systems.  Note that orbital decay with $P_{\rm orb}/ \left| {\rm d} P_{\rm orb}/ {\rm d}t \right|=3.2$~Myr is compatible with observations for the Jupiter mass planet WASP-12b, which has an orbital period of 1.09~day (\citealt{Patra2017}, see also \citealt{Maciejewski2016}).  This would correspond to $t_{\rm orb} \simeq 5$~Myr, which is very close to the value of 6~Myr we obtain here.  

{\em Energy dissipation and inflated radii:}  Some giant extrasolar planets are observed to have an anomalously large radius. Starting with the work by \citet{Bodenheimer2001},  tidal dissipation has been proposed as a mean to inflate those planets. However, subsequent studies have found that, even if the rate of tidal dissipation is adjusted such as to account for the circularization of late--type binaries, it is not large enough to account for the inflated radius of hot Jupiters \citep{Leconte2010}.  
As planets synchronize relatively fast, energy can only be dissipated by tides raised in the planet by the star if the orbit retains some eccentricity.  In this case, the energy dissipation rate is given by equation~(\ref{eq:dEdtesync}), and is  proportional to $e^2$.  As orbits with periods smaller than about 5~days circularize on timescales of a few Gyr, eccentricities are very small, as confirmed by observations, which limits the rate of energy dissipation.  

\begin{figure*}
    \centering
   \includegraphics[width=\columnwidth,angle=270]{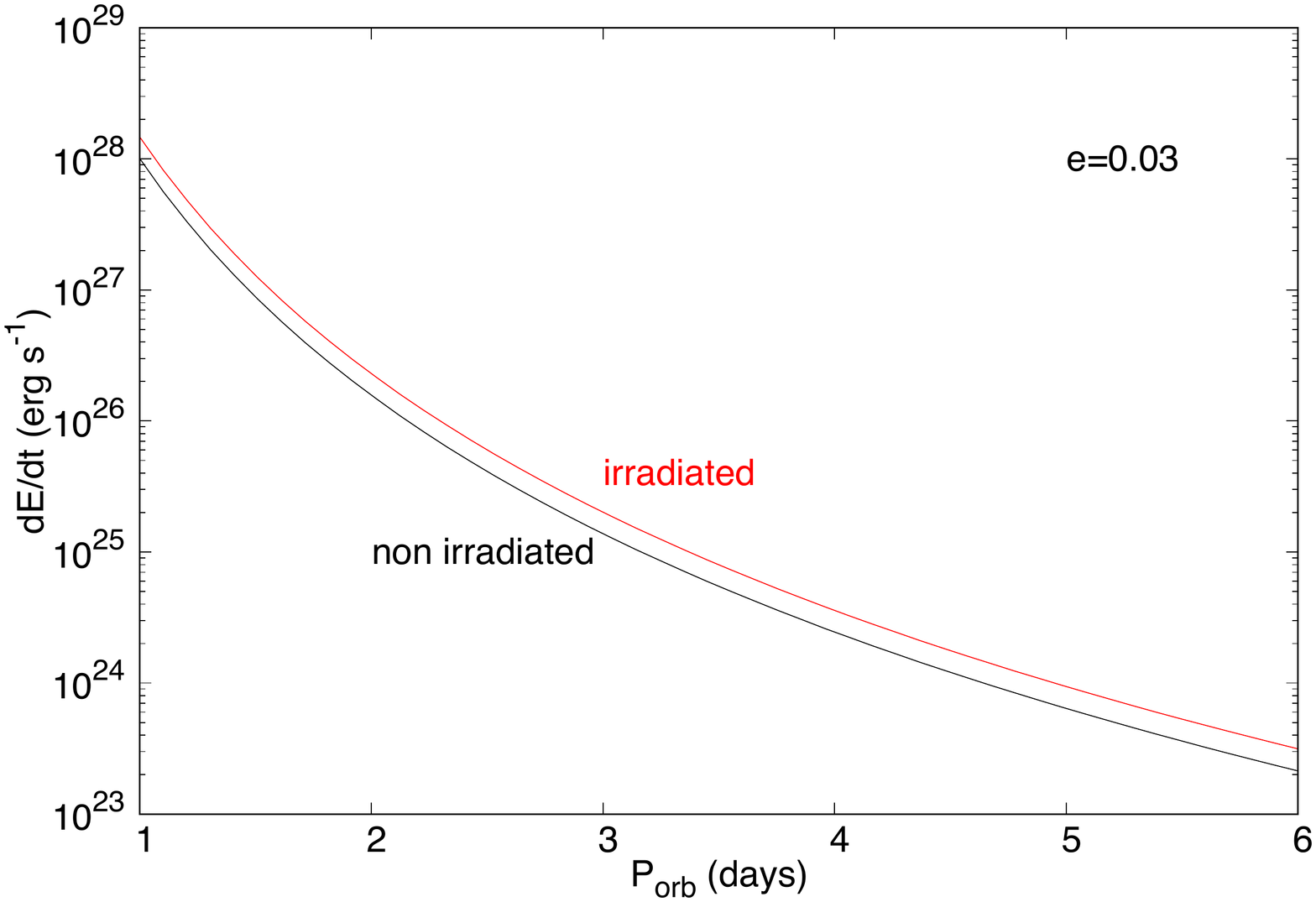}
    \caption{Rate of energy dissipation $\left( {\rm d}E / {\rm d}t \right)_{\rm e,sync}$ (calculated from eq.~[\ref{eq:dEdtesync}]) in erg~s$^{-1}$ using a logarithmic scale {\em versus} orbital period in days for 
    a Jupiter model (black curve) and an irradiated Jupiter model (red curve) in which tides are raised by a 1~M$_{\sun}$ star, for an eccentricity $e=0.03$.  
      }
    \label{fig7}
\end{figure*}

Figure~\ref{fig7} shows the rate of energy dissipation $\left( {\rm d}E / {\rm d}t \right)_{\rm e,sync}$ calculated from equation~(\ref{eq:dEdtesync}) for 
both a standard Jupiter model and an irradiated Jupiter model  in which tides are raised by a 1~M$_{\sun}$ star, assuming an eccentricity $e=0.03$.  This is an upper limit for most of the systems in which an inflated radius is present   \citep{Jackson2008}.  To explain the  inflated radii which are observed to be  between 1.1 and 1.5~$R_{\rm J}$ for a large number of hot Jupiters, a heating rate between $10^{25}$ and $10^{28}$~ergs~s$^{-1}$ is needed (\citealt{Miller2009, Bodenheimer2003}).  These are the values we obtain only for orbital periods smaller than 3~days.
Therefore, our results confirm that tidal dissipation alone cannot explain the inflated radius of most hot Jupiters. 

\section{Discussion and conclusion}
\label{sec:Discussion}

The models for Jupiter  in figure~\ref{fig2_modeljupiter} and Saturn in figure~\ref{fig2_modelsaturn} show that the convective timescales in the envelope of the planets are much larger than the tidal periods of interest.  Therefore, the timescales of convection and  the tides are well separated,  which validates the analysis carried out in section~\ref{sec:conservation}.  This analysis shows from first principles that the rate $D_R$ at which energy per unit mass is exchanged between the tides and the convective flow  {\em via} the Reynolds stress
is  given by equation~(\ref{eq:DR}), where ${\bf u}'$ is the velocity of the tides and ${\bf V}$  the velocity of the convective flow.  This is in contrast to the standard approach which has been used in previous studies, and which identifies the mean flow and the fluctuations based on the spatial scales on which they vary, rather than on the timescales, therefore interchanging the role of the tidal and convective velocities in equation~(\ref{eq:DR}).  
Figure~\ref{fig2_modeljupiter} also shows that  the diffusion approximation, which has been used to express the convective Reynolds stress as a turbulent viscosity,  is not self--consistent, even in the modified form which accounts for a suppression of dissipation at long turnover timescales, as the scale of the convective eddies $l_m$  is large or comparable to the radius $r$ in a large part of the atmosphere.   Below $r=0.5R_{\rm J}$, $r/l_m <1$, and $r/l_m$ reaches 5 only at $r=0.8R_{\rm J}$.    
Similar results apply to models of Saturn.
For the Sun, as seen in figure~\ref{fig1_modelsun},  convective timescales are  large compared to tidal periods of interest in the inner parts of the convective envelope, and $l_m$ is only moderately smaller than $r$ there.   The non--locality of convection in the Sun has of course been known for a long time, and non--local theories of convection have been proposed \citep{Spiegel1963, Unno1969, Ulrich1970, Xiong1979}.

Note that, although we are arguing that mixing length theory does not apply in the envelopes of giant planets and the Sun, we have used the convective velocities $V$ and timescales $t_{\rm conv}$ from models based on this approximation in section~\ref{sec:applications}.  However,  for slow rotators like the Sun, the orders of magnitude of $V$ and $t_{\rm conv}$ (and hence $l_m= V t_{\rm conv}$) do not actually depend on the details of the model, and could be obtained directly from dimensional analysis by matching the convective flux of energy to the observed flux, as done at the beginning of section~\ref{sec:conservation}.   That being said, it is worth keeping in mind that the convective velocities required to transport the energy radiated by the Sun seem to be larger than those needed to establish  differential rotation and those inferred by observations \citep{OMara2016}. In fast rotators, it has been proposed that $t_{\rm conv} \propto  {\rm Ro}^{2/5}$, where  Ro  is the Rossby number based on convective velocities in the absence of rotation (\citealt{Stevenson1979}, \citealt{Barker2014}, \citealt{Gastine2016}).   In giant planets, ${\rm Ro} \sim 10^{-5}$--$10^{-4}$,  which yields convective timescales about two orders of magnitude smaller than those used here.  This would correspond to much smaller values of $Q$, as shown by equation~(\ref{eq:Qapprox}).  It is not clear however whether models including such a dramatic change in the convective timescales would agree with observations.   The studies leading to this scaling are in essence an extension of the  mixing length theory to rotating systems, which may not be a good description of convection in fast rotating bodies.

The formal derivation of the rate $D_R$ at which  energy is exchanged between  the tides and the convective flow with large turnover timescale  is a robust result.  However, calculating this term specifically  in the envelopes of the Sun or giant planets would require knowing the velocity of the convective flow there, which can only  be achieved by numerical simulations.    
A positive $D_R$ would mean that energy is locally transferred from the tides to the convective flow, whereas a negative $D_R$ would mean that energy is fed to the tides.  It may even be that $D_R$ changes sign depending on location.   However, circularization of 
 late--type binaries and the orbital evolution of the moons of Jupiter and Saturn require tides to dissipate in the convective envelopes of stars and giant planets.   We have accordingly calculated the evolution timescales for these systems assuming $D_R$ to be positive everywhere in the interiors of stars and planets, which yields maximal energy dissipation, and investigated whether this led to timescales in agreement with observations.  The timescales we obtain match very well the observations for Jupiter and PMS binaries, and also for Saturn when adopting recent models in which the lengthscale over which the convective velocity varies is smaller than that given by standard mixing length theory \citep{Vazan2016}.  Such a reduction in this lengthscale has been suggested for giant planets by \citet{Leconte2012}.  It is also consistent with  studies which find that, in rotating bodies, the mixing length is reduced by a factor equal to
 ${\rm Ro}'$ \citep{Vasil2020} or $2 {\rm Ro}^{3/5}$ (\citealt{Stevenson1979}, \citealt{Barker2014}, \citealt{Currie2020}), where  ${\rm Ro}'$ is the Rossby number based on convective velocities in the presence of rotation.
 This is because the Taylor--Proudman theorem favours rotation along cylinders centered on the rotation axis, therefore reducing the scale of the flow perpendicular to the axis.   However,  as pointed out above, it is not clear whether mixing length theory applies in the presence of fast rotation.
 
 For Jupiter and Saturn, an additional source of tidal dissipation may be provided by gravity modes which are excited in stably stratified layers.  Such layers have recently been shown to  be compatible with Juno's gravity measurements  of Jupiter \citep{Wahl2017}.  For Saturn, stable layers are predicted by recent models \citep{Vazan2016} and  also by the analysis of density waves within the rings \citep{Fuller2014}.   Resonance locking between satellites and gravity modes in evolving planets has been proposed as an explanation for the low $Q$ values of both Jupiter and Saturn \citep{Fuller2016}.
 
 The fact that our results do not match the observations for late--type binaries, whereas they yield good agreement for bodies which are fully convective, is indicative  that tidal dissipation in solar type stars may be due to the shear present at the base of the convective envelope, where convective velocities go to zero rather abruptly, or in the tachocline, where the rotational velocity has a strong radial gradient.   The component of the Reynolds stress which couples to this shear is $\left< u'_r u'_{\varphi} \right>$.  This is zero when there is no dissipation, as $u'_r$ and $u'_{\varphi}$ are $\pi/2$ out of phase in that case, but this could become significant in regions where dissipation is large, as this introduces an additional phase shift  (e.g., \citealt{Bunting2019}).
 
 Dissipation of inertial waves in the convective envelope has also been considered as a possible explanation for  the observed  circularization periods. These waves are excited when the tidal frequency in the frame of the fluid, $\left| n \omega_{\rm orb} -m \Omega \right| $, where $\Omega$ is the (uniform) angular velocity of the star, is smaller than $2 \Omega$.    As synchronization  of the stars happens much more rapidly than circularization,  $\omega_{\rm orb} = \Omega $ during most of the circularization phase and inertial waves are excited by the terms in the tidal potential which are  first order in eccentricity, which correspond to $n-m= \pm 1$ \citep{Ogilvie2014}.  
\citet{Ogilvie2007}, and more recently \citet{Barker2020}, have shown that the rate of energy dissipation of these waves in the convective zone is much larger than that of equilibrium tides when mixing length theory is used for those.    \citet{Barker2020} obtains a circularization timescale of 1~Gyr for an orbital period of  7~days (this result corresponds to  dissipation in a single solar-mass star,  but it would hardly change   if tides in both stars were taken into account).  Although this process is slightly more efficient than the one discussed here, it still does not account for the observed circularization periods.  
 
The good agreement between our results and the observations for fully convective bodies  is of course not by itself  a proof that $D_R>0$, but  indicates that the model presented here is a route worth exploring further.  
It also suggests that there may be a mechanism by which the convective flow re-arranges itself to always extract energy from the tides.   It has been known for some time that the interaction of rotation with convection in the envelope of the Sun produces large--scale axisymmetric flows that extend in the entire convective envelope.    The most striking feature of these flows is the differential rotation in the latitudinal direction, which makes the poles rotate 30\% slower than the equator all the way through the convective zone.  Global torsional oscillations  in the longitudinal direction \citep{Howe2018} and a large scale meridional flow have also been observed.  The meridional flow  involves motions in both the latitudinal and radial directions and takes the form of a single cell in each hemisphere of the Sun \citep{Gizon2020}.   
Numerical simulations of this meridional flow show that, like differential rotation,  it is established  by angular momentum transport resulting from the convective Reynolds stress  in the presence of rotation
(e.g., \citealt{Featherstone2015, Hotta2015}).
In addition,   numerical simulations show that rotation inhibits radial downdrafts near the equator and produces prominent columnar structures aligned with the star rotation axis \citep{Featherstone2015}, consistent with the Taylor--Proudman theorem.  

Although the velocities associated with the large scale flows are  much smaller than the  convective velocities, and would therefore not themselves provide a large shear the tidal Reynolds stress could couple to, 
these results suggest 
that the structure of the convective flow in a rotating body is very different from the simple standard picture, where fluid elements move up and down resulting in a convective velocity which averages to zero spatially.   

In Jupiter, as already mentioned in section~\ref{sec:Jupiter}, it has been found that differential rotation is limited to the upper 4\% or so of the atmosphere.  Therefore, convection in this planet may not  generate large scale flows deeper in the atmopshere.  This would however not be inconsistent with our results, as we have found that the upper 10\% of Jupiter's atmosphere could account for its tidal dissipation factor. 

Whether the interaction between convection, rotation and the tides can  produce the convective velocity gradients required for $D_R$ to be positive  could be tested by measuring this term in numerical simulations.   It would also be interesting to know how the circularization period of late--type binaries varies with stellar rotation: if large scale flows in convective envelopes are important in providing the right gradient of convective velocity to make $D_R>0$, then tidal dissipation should be more efficient in more rapidly rotating stars, in which more global structures develop \citep{Featherstone2015}.

\section*{Acknowledgements}

I am very grateful to  Isabelle Baraffe for providing models of Jupiter, Saturn, an irradiated Jupiter and the Sun,
to Ravit Helled, Allona Vazan, Yamila Miguel and Tristan Guillot for sharing their latest models of Saturn, and for their patience in answering all my questions and requests.  
I also thank
 Steven Balbus for encouragements and very stimulating discussions, Gilles Chabrier for very useful insight into models of giant planets, Jeremy Goodman, Henrik Latter, Gordon Ogilvie and John Papaloizou for feedback on an early version of this paper, and  Robert Mathieu for observational updates on the most recent circularization periods for late--type binaries.   
 Finally, I thank the referee, Adrian Barker, for a very thorough and constructive review which has  improved the manuscript. 
 This work used the {\em Modules for Experiments
in Stellar Astrophysics} (MESA) code available from {\em mesa.sourceforge.net}.   

\section*{Data availability}

No new data were generated or analysed in support of this research.










\appendix

\section{Energy conservation in spherical coordinates}
\label{appendix}

We consider a spherical coordinate system $(r, \theta , \varphi)$ centered on the star and denote the associated unit vectors  ${\bf e}_r , {\bf e}_{\theta}, {\bf e}_{\varphi}$. The equation for conservation of energy of the mean flow is obtained as described in section~\ref{sec:conservation}.  In spherical coordinates, neglecting viscous dissipation,  this yields:
\begin{multline}
\frac{\partial }{\partial t } \left( \frac{1}{2} V^2 \right) + \left( {\bf V} \cdot \nab \right) \left( \frac{1}{2} V^2 \right) =  \\  - \frac{\partial}{\partial x_j} \left( V_i \left< u'_j u'_i \right> \right) -   \nab \cdot \left(   \frac{1}{\rho} P {\bf V} \right) + \frac{1}{\rho} {\bf f} \cdot {\bf V} +D_R,
\end{multline}
with:
\begin{equation}
V^2=V^2_r + V^2_{\theta} + V^2_{\varphi},
\end{equation}
\begin{equation}
 {\bf V} \cdot \nab = V_r \frac{\partial }{\partial r}+ \frac{V_{\theta}}{r} \frac{\partial }{\partial \theta } + \frac{V_{\varphi}}{r \sin \theta } \frac{\partial }{\partial \varphi} ,
\end{equation}
\begin{multline}
\frac{\partial}{\partial x_j} \left( V_i \left< u'_j u'_i \right> \right) = \\
\frac{\partial }{\partial r} \left( V_r \left< u'^2_{r} \right> + V_{\theta} \left< u'_r u'_{\theta} \right> + V_{\varphi} \left< u'_r u'_{\varphi} \right> \right) \\ 
+ \frac{1}{r} \frac{\partial }{\partial \theta} \left( V_r   \left<  u'_{\theta} u'_r \right> + V_{\theta} \left< u'^2_{\theta} \right> + V_{\varphi} \left<  u'_{\theta} u'_{\varphi} \right>  \right) \\
+ \frac{1}{r \sin \theta} \frac{\partial }{\partial \varphi} \left( V_r   \left<  u'_{\varphi} u'_r \right> + V_{\theta} \left< u'_{\varphi} u'_{\theta} \right> + V_{\varphi} \left<  u'^2_{\varphi} \right>  \right) \\
+ V_r \left( \frac{2  }{r} \left< u'^2_r \right> + \frac{ \cot \theta  }{r}   \left< u'_r u'_{\theta} \right> \right) + V_{\theta} \left( \frac{2  }{r} \left< u'_r u'_{\theta} \right> + \frac{ \cot \theta  }{r}  \left< u'^2_{\theta} \right> \right) \\
+ V_{\varphi} \left( \frac{2 }{r}  \left< u'_r u'_{\varphi} \right>  + \frac{ \cot \theta  }{r}  \left< u'_{\theta} u'_{\varphi} \right> \right) ,
\end{multline}
\begin{multline}
\nab \cdot \left(   \frac{1 }{ \rho} P {\bf V}  \right) = \frac{\partial}{\partial r} \left( \frac{V_r P}{ \rho}  \right) +  \frac{1}{r} \frac{\partial}{\partial \theta } \left( 
\frac{V_{\theta } P}{ \rho} \right) + \frac{1}{r \sin \theta } \frac{\partial}{\partial \varphi } \left( \frac{V_{\varphi  } P}{\rho} \right) \\ + \frac{2}{r } \frac{ V_r P}{\rho} + \frac{ \cot \theta }{r} \frac{ V_{\theta} P }{ \rho}, 
\end{multline}
\begin{multline}
D_R=  \left< u'^2_r \right>  \frac{\partial V_r}{\partial r} + \left< u'^2_{\theta} \right> \left( \frac{1}{r} V_r + \frac{1}{r} \frac{\partial V_{\theta}}{\partial \theta } \right)
\\ + \left< u'^2_{\varphi} \right> \left( \frac{1}{r} V_r   + \frac{ \cot \theta }{r} V_{\theta} + \frac{1}{r \sin \theta } \frac{\partial V_{\varphi }}{\partial \varphi } \right) \\
+  \left< u'_r u'_{\theta }  \right> \left( r \frac{\partial }{\partial r} \left( \frac{V_{ \theta }}{r} \right) + \frac{1}{r} \frac{\partial V_r }{\partial \theta } \right) \\
+  \left< u'_r u'_{\varphi }  \right> \left( r \frac{\partial }{\partial r} \left( \frac{V_{ \varphi }}{r} \right) + \frac{1}{r \sin \theta } \frac{\partial V_r }{\partial \varphi } \right) \\ + \left< u'_{\theta } u'_{\varphi}  \right> \left( \frac{ \sin \theta}{r} \frac{\partial }{\partial \theta} \left( \frac{V_{\varphi}}{\sin \theta} \right) + \frac{1}{r \sin \theta} \frac{\partial V_{\theta}}{\partial \varphi } \right)
\label{eq:DRspher}
\end{multline}
Locally, we can define a Cartesian coordinate system $(x,y,z)$ such that the $x$, $y$ and $z$--axes are along ${\bf e}_{\theta}$, ${\bf e}_{\varphi}$ and ${\bf e}_r$, respectively.  Therefore, ${\rm d} x= r {\rm d} \theta$, ${\rm d} y= r \sin \theta  {\rm d} \varphi$ and ${\rm d} z=  {\rm d} r$.  If the curvature is locally negligible (i.e., $r \gg \left| {\rm d}x \right| , \left| {\rm d}y \right| , \left| {\rm d}z \right| $), then $D_R$ reduces to:
\begin{multline}
D_R= \left< u'^2_z \right>  \frac{\partial V_z}{\partial z} + \left< u'^2_{x} \right> \frac{\partial V_{x}}{\partial x} 
+ \left< u'^2_{y} \right>  \frac{\partial V_{y }}{\partial y} \\
+  \left< u'_z u'_{x }  \right> \left( \frac{\partial V_x}{\partial z} + \frac{\partial V_z }{\partial x} \right) 
+  \left< u'_z u'_{y }  \right> \left( \frac{\partial V_y}{\partial z} +  \frac{\partial V_z }{\partial y } \right)  \\ + \left< u'_{x } u'_{y}  \right> \left(  \frac{\partial V_y }{\partial x} +  \frac{\partial V_{x}}{\partial y } \right),
\end{multline}
so that we recover  expression~(\ref{eq:DR})  in Cartesian coordinates.  

\bsp	
\label{lastpage}
\end{document}